\documentclass[12pt]{article} 
\usepackage[usenames,dvipsnames]{color}

\usepackage{graphicx}
\usepackage{amsthm}
\usepackage{graphics}
\usepackage{amsmath,,amssymb}
\usepackage{algorithmic}
\usepackage{algorithm}
\usepackage{caption}
\usepackage{subcaption}
\usepackage{pstricks, pst-node}
\usepackage{tikz}
\usepackage[linewidth=1pt]{mdframed}
\usepackage{multirow}
\usepackage[round,colon,authoryear]{natbib}

\newtheorem{theorem}{Theorem}

\newtheorem{corollary}{Corollary}

\def\RR{\mathbb{R}}
\def\EE{\mathbb{E}}
\def\II{\mathbb{I}}
\def\PP{\mathbb{P}}
\def\bc{\mathbf{c}}
\def\bx{\mathbf{x}}
\def\bX{\mathbf{X}}
\def\bZ{\mathbf{Z}}
\def\bu{\mathbf{u}}
\def\bv{\mathbf{v}}
\def\bw{\mathbf{w}}

\def\calU{\mathcal{U}}

\def\tr{\mathrm{trace}}
\def\one{\mathbf{1}}
\def\zero{\mathbf{0}}

\DeclareMathOperator*{\argmax}{arg\,max}

\DeclareMathOperator*{\esssup}{ess\,sup}
\def\bmu{\mbox{\boldmath $\mu$}}
\def\beps{\mbox{\boldmath $\epsilon$}}

\begin{document}
\title{Combined Hypothesis Testing on Graphs with Applications to Gene Set Enrichment Analysis$^\ast$}

\date{(\today)}

\author{Shulei Wang and Ming Yuan$^\dag$\\
Morgridge Institute for Research and University of Wisconsin-Madison}

\footnotetext[1]{
Research supported in part by NSF FRG Grant DMS-1265202, and NIH Grant 1-U54AI117924-01.}
\footnotetext[2]{Address for Correspondence: Department of Statistics, University of Wisconsin-Madison, 1300 University Avenue, Madison, WI 53706.}
\maketitle

\begin{abstract}
Motivated by gene set enrichment analysis, we investigate the problem of combined hypothesis testing on a graph. We introduce a general framework to effectively use the structural information of the underlying graph when testing multivariate means. A new testing procedure is proposed within this framework. We show that the test is optimal in that it can consistently detect departure from the collective null at a rate that no other test could improve, for almost all graphs. We also provide general performance bounds for the proposed test under any specific graph, and illustrate their utility through several common types of graphs. Numerical experiments are presented to further demonstrate the merits of our approach.
\end{abstract}

\newpage
\section{Introduction}
Combined hypothesis testing arises naturally in many modern statistical applications. See Chapter 9 of \cite{Efron2013} for further discussions. The most notable example is the so-called gene set enrichment analysis. See, e.g., \cite{Mootha2003}, \cite{Subramanian2005}, \cite{Tian2005}, \cite{EfronTibshirani2007}, \cite{GoemanBuhlmann2007}, \cite{JiangGentleman2007}, \cite{Newton2007}, and \cite{AckermannStrimmer2009} among many others. It is motivated by the observation that many complex diseases are manifested through modest regulation in a set of related genes rather than a strong effect on a single gene. While statistical testing of the regulatory effect on a particular gene may be inconclusive, the collective effect on a set of genes can oftentimes be clearly identified through combined hypothesis testing. The results produced by gene set enrichment analysis could therefore be more reliable and biologically meaningful when compared with those based on a single gene. Given its importance in statistical genomics, gene set enrichment analysis has attracted much attention in recent years and numerous approaches have been proposed. See \cite{Maciejewski2013} and \cite{NewtonWang2015} for a couple of recent surveys on existing techniques. 

Most statistical methods for gene set enrichment analysis proceed in two steps. One first computes for each gene a local statistic, i.e., for testing if a gene is differentially expressed between multiple biological conditions. For concreteness, denote by $V$ the index set of a particular collection of related genes, and $x_v$ the $z$-score associated with the $v$th gene so that
$$
x_v=\mu_v+\epsilon_v,\qquad v\in V,
$$
where $\epsilon_v\sim N(0,1)$. In the second step, we consider testing a combined null hypothesis that there is no effect on the gene set, that is
\begin{equation}
\label{eq:null}
H_0: \mu_v=0,\qquad \forall v\in V,
\end{equation}
against an overall effect:
\begin{equation}
\label{eq:altall}
H_a: \bmu=(\mu_v)_{v\in V}\neq {\bf 0}.
\end{equation}
See, e.g., \cite{Efron2013}.

A rich source of information often neglected in these analyses is the fact that a gene set is typically taken from a certain biological pathway, be it a metabolic pathway or a signaling pathway, describing a series of biochemical and molecular steps towards a specific biological function. Many of the known pathways are now readily accessible through several well-curated databases such as Gene Ontology \citep{Ashburner2000}, KEGG \citep{KanehisaGoto2000}, or Pathguide \citep{BaderCarySander2006}. A pathway can be conveniently represented by a graph $G=(V,E)$ where each node $v\in V$ corresponds to a gene, and an edge $(v_1,v_2)\in E$ between a pair of nodes indicates direct interactions between them. It is, however, largely unknown to what extent such pathway information could be utilized in gene set enrichment analysis. The main goal of this article is to address this issue, and develop a principled and effective way to take advantage of such structural information for gene set enrichment analysis in particular and combined hypothesis testing in general.

More specifically, we introduce a hierarchy among all possible effects based on their level of smoothness with respect to the underlying pathway, and argue that the difficulty in testing against a particular effect $\bmu(\neq 0)$ depends critically on its smoothness in that ``smoother'' effects are ``easier'' to detect. We note, that unlike functions defined over a continuous domain, smoothness is an innocuous concept here because any $\bmu\in \RR^{|V|}$ can be associated with a finite smoothness index. This framework allows us to exploit the fact that, in many applications, it is plausible that a putative effect is sufficiently ``smooth''. But at the first glance, such an observation may have little practical implication because even if the effect is indeed smooth, one rarely knows how smooth it might be. We show here that despite the absence of such knowledge regarding $\bmu$'s smoothness, it is still possible to develop an agnostic test that automatically {\it adapts} to it. In particular, we develop an easily implementable adaptive testing procedure whose power increases automatically with the smoothness of the unknown $\bmu$.

To demonstrate the merits of the proposed paradigm and test, we study its asymptotic properties from two different and complementary aspects: an average-case analysis based on Erd\"os-R\'enyi model; and a general analysis that applies to any specific type of graphs. The former analysis shows that among all graphs of $n$ nodes, with the exception of a vanishing proportion of graphs under Erd\"os-R\'enyi model, the proposed test is minimax optimal for any level of smoothness in that one cannot do better over all effects at the same level of smoothness even if we know in advance how smooth they are. In addition, we derive a generally applicable performance bound for our test and illustrate through several fundamental types of graph its utility and optimality.

Although we focus our discussion primarily in the context of gene set enrichment analysis, it is worth noting that the methodology and theory we developed here may also be useful in many other applications. For example, one may be interested in performing combined hypothesis testing over locations within a particular region of the brain, as a means to identifying areas that can be associated with certain activities. See, e.g., \cite{ChungHansonPollak2016}. In these situations, it is plausible that an overall effect is smooth with respect to the brain surface manifold. This can be translated into smoothness with respect to nearest neighbor graphs underlying these locations. The techniques developed here can then be employed in these applications.

The rest of the paper is organized as follows. We first introduce the general framework of our treatment and the proposed test in Section \ref{sec:meth}. In Section \ref{sec:ave} we investigate the properties of the proposed tests under the Erd\"os-R\'enyi model to gain insights into their operating characteristics as well as the effect of smoothness on the detectability of a particular effect. Section \ref{sec:spe} provides general performance bounds for our tests, and their applications to several common types of graphs. Numerical experiments are presented in Section \ref{sec:sim} to further demonstrate the merits of the proposed methodology. %We conclude with a few remarks in Section \ref{sec:conclude}. 
All proofs are relegated to Section \ref{sec:proof}.

\section{Methodology}
\label{sec:meth}

A natural approach to testing $H_0$ against $H_a$ given by Equations (\ref{eq:null}) and (\ref{eq:altall}) respectively is a $\chi^2$-test based on the statistic
$$\|\bX\|^2:=\sum_v X_v^2$$
where $\bX=(X_v)_{v\in V}$. It is clear that under $H_0$, $\|\bX\|^2$ follows a $\chi^2_{|V|}$ distribution, so that a $\alpha$-level test rejects $H_0$ if and only if $\|\bX\|^2$ exceeds the $(1-\alpha)$ quantile, denoted by $\chi^2_{|V|,\alpha}$, of $\chi^2_{|V|}$ distribution. Here $|\cdot|$ denotes the cardinality of a set. On the other hand, under the alternative hypothesis, $\bX\sim N(\bmu,I)$ so that $\|\bX\|^2$ follows a non-central $\chi^2_{|V|}(\|\bmu\|^2)$ distribution. Denote by $\varphi_{\chi^2,\alpha}$ the $\alpha$-level $\chi^2$-test. Hereafter, we shall omit the subscript $\alpha$ and write $\varphi_{\chi^2}$ for brevity, when no confusion occurs. It is clear that the Type II error of $\varphi_{\chi^2}$ is given by
$$
\beta(\varphi_{\chi^2}; \bmu):=\PP_{\bX\sim N(\bmu,I)}\left\{\|\bX\|^2> \chi^2_{|V|,\alpha}\right\}.
$$
It is not hard to see that $\beta(\varphi_{\chi^2}; \bmu)$ goes to zero as soon as $\|\bmu\|^2\gg |V|^{1/2}$, where $a_{V}\gg b_{V}$ means $a_{V}/b_{V}\to \infty$ as $|V|\to \infty$. In other words, $\varphi_{\chi^2}$ can consistently detect all effects $\bmu$ such that
\begin{equation}
\label{eq:bdchi2}
\|\bmu\|^2\gg |V|^{1/2}.
\end{equation}
Furthermore, it is well known that the $\chi^2$-test is minimax optimal in testing $H_0$ against $H_a$ in that the detection boundary given by (\ref{eq:bdchi2}) cannot be improved. More precisely, there exists a constant $c>0$ such that for any $\alpha$-level ($0<\alpha<1$) test $\Psi$ based on $\bX$,
$$
\liminf_{|V|\to \infty}\sup_{\bmu: \|\bmu\|^2\ge c|V|^{1/2}}\beta(\Psi,\bmu)>0.
$$
See, e.g., \cite{IngsterSuslina2003} for further discussions.

Despite the minimax optimality of $\chi^2$-test, there is also ample empirical evidence that other tests, such as $z$-test, may work better in some situations. This is because the optimality of $\chi^2$-test is in the minimax sense, and therefore based on the worst-case performance. Although the minimax optimality suggests that no test could do better than $\varphi_{\chi^2}$ over all effects $\bmu\in \RR^{|V|}\setminus \{{\bf 0}\}$, it does not necessarily preclude improvements over subsets of $\RR^{|V|}$. For example, if $\bmu\propto \one$, where ${\bf 1}$ is the vector of ones, then $z$-test is a more powerful test than $\chi^2$-test and it can detect any $\bmu$ of this form as long as $\|\bmu\|\to \infty$. In fact, $z$-test is the most powerful test in this situation by Neyman-Pearson Lemma. Unfortunately, such an improvement over $\chi^2$-test comes at a hefty price -- $z$-test is powerless in testing against any effect $\bmu$ such that $\bmu^\top\one=0$ in that
$$
\beta(\varphi_z;\bmu)=1-\alpha
$$
where $\varphi_z$ denotes the $\alpha$-level $z$-test. 

This naturally brings about the question of whether or not the strengths of $z$-test and $\chi^2$-test could be combined. We show that this indeed is possible, and develop a test that is just as powerful as the $\chi^2$-test in the absence of any information regarding a putative effect, but could be as powerful as the $z$-test when the effect is indeed a constant. More generally, the test could be substantially more powerful than the $\chi^2$-test depending on the smoothness of the unknown effect with respect to the graph $G=(V,E)$. This is of particular interest here because in many applications of gene set enrichment analysis, it is plausible that the effect $\bmu$ is of certain level of smoothness with respect to the graph $G$. Our framework here is largely inspired by the pioneering work of \cite{Ingster1993} on nonparametric testing. See also \cite{IngsterSuslina2003}.

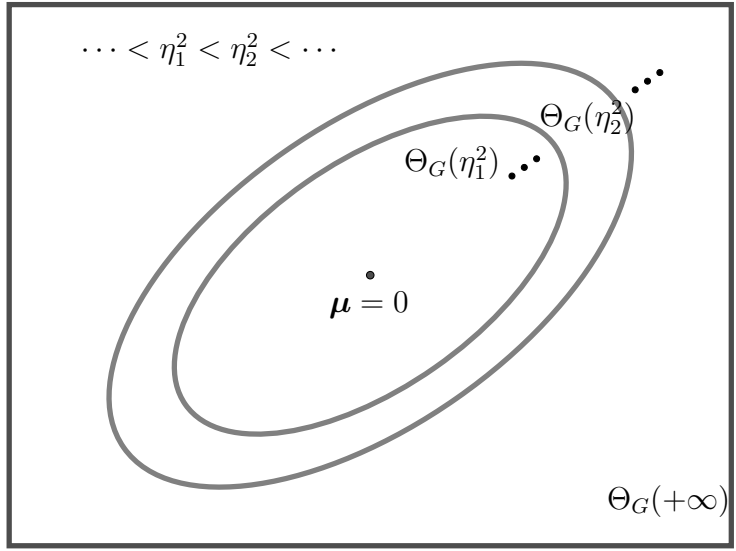
\begin{figure}[htbp]
\begin{center}
\begin{tikzpicture}
 \begin{scope}
       \fill[white,fill opacity=0.8] (-4.8,-3.6) rectangle (4.8,3.6); 
       \path[draw,line width=2pt,black!70!white] (-4.8,-3.6) -- (-4.8,3.6)--(4.8,3.6)--(4.8,-3.6)--cycle;
       \draw[rotate=35,line width=2pt,black!50!white] (0,0) ellipse (3 and 1.5);
       \draw[rotate=35,line width=2pt,black!50!white] (0,0) ellipse (4 and 2);
       %\draw[rotate=35,line width=2pt,black!50!white] (0,0) ellipse (5 and 2.5);
       
       \draw[rotate=35,line width=0pt,fill=black] (4.3,0) ellipse (0.04 and 0.04);
       \draw[rotate=35,line width=0pt,fill=black] (4.5,0) ellipse (0.04 and 0.04);
       \draw[rotate=35,line width=0pt,fill=black] (4.7,0) ellipse (0.04 and 0.04);
       \draw[rotate=35,line width=0pt,fill=black] (2.3,0) ellipse (0.04 and 0.04);
       \draw[rotate=35,line width=0pt,fill=black] (2.5,0) ellipse (0.04 and 0.04);
       \draw[rotate=35,line width=0pt,fill=black] (2.7,0) ellipse (0.04 and 0.04);
       \draw[fill=black!70!white] (0,0) ellipse (0.05 and 0.05);
 \end{scope}
 \draw[thick,black](-0.7,-0.4)node[right]{$\bmu=0$};
 \draw[thick,black](0.3,1.5)node[right]{$\Theta_G(\eta_1^2)$};
 \draw[thick,black](2.1,2.1)node[right]{$\Theta_G(\eta_2^2)$};
  \draw[thick,black](3,-3)node[right]{$\Theta_G(+\infty)$};
  \draw[thick,black](-4,3)node[right]{$\cdots<\eta_1^2<\eta_2^2<\cdots$};
\end{tikzpicture}
\end{center}
\caption{Smoothness creates an hierarchy among all effects in $\RR^{|V|}$.}
\label{fig:Theta}
\end{figure}

Recall that the Laplacian matrix of $G$ is given by
$$
L(G)=D(G)-A(G)
$$
where $D(G)$ and $A(G)$ are its degree matrix and adjacency matrix respectively. To fix ideas, we shall focus on unweighted and undirected graphs, although our treatment can also be applied to more general, e.g., weighted or directed, graphs. For an unweighted and undirected graph $G$, the adjacency matrix $A(G)$ is a symmetric matrix whose $(v,v')$ entry is one if $(v,v')\in E$ and zero otherwise, and the degree matrix $D(G)$ is a diagonal matrix whose $v$th diagonal entry is the degree of node $v$. It is clear that
\begin{equation}
\label{eq:smoothind}
\bmu^\top L(G) \bmu=\sum_{(v_1,v_2)\in E}(\mu_{v_1}-\mu_{v_2})^2,
\end{equation}
so that it measures the smoothness of $\bmu$ with respect to $G$. The smoothness of $\bmu$ with respect to graph $G$ allows us to create a hierarchy in $\RR^{|V|}$. More specifically, for an arbitrary $\eta^2\ge 0$, denote by $\Theta_G(\eta^2)$ the collection of all effects whose smooth index as defined by (\ref{eq:smoothind}) is at most $\eta^2$, that is,
\begin{equation}
\label{eq:defTheta}
\Theta_G(\eta^2)=\{\bmu\in \RR^{n}: \bmu^\top L(G)\bmu\le \eta^2\}.
\end{equation}
For brevity, we shall omit the subscript $G$ in what follows when no confusion occurs. Obviously, the smaller $\eta^2$ is, the smaller $\Theta(\eta^2)$ is, as illustrated in Figure \ref{fig:Theta}. Thus, it is natural to expect it to be easier to detect effects from $\Theta(\eta^2)$ for smaller $\eta^2$s. In particular, since $\chi^2$-test is optimal for testing against an arbitrary effect $\bmu\in \Theta(+\infty)=\RR^{|V|}$, we might expect to be able to improve it over $\Theta(\eta^2)$ for any finite $\eta^2$. It turns out, however, not to be the case.

\begin{theorem}
\label{th:s2optimal}
Let $\eta_{\max}^2(G)=|V|^{-1/2}\tr(L(G))$ where $L(G)$ is the Laplacian of a graph $G=(V,E)$. Then for any $\eta^2=\Omega(\eta_{\max}^2(G))$ there exists a constant $c>0$ such that for any $\alpha$-level ($0<\alpha<1$) test $\Psi$,
$$
\liminf_{|V|\to\infty}\sup_{\bmu\in \Theta_G(\eta^2): \|\bmu\|^2\ge c|V|^{1/2}}\beta(\Psi;\bmu)>0.
$$
\end{theorem}

Hereafter, we write $a_{V}=\Omega(b_{V})$ if $b_{V}=O(a_{V})$. Theorem \ref{th:s2optimal}, together with the fact that $\chi^2$-test is consistent for any $\bmu\in \RR^{|V|}$ such that $\|\bmu\|^2\gg |V|^{1/2}$, suggests that one cannot improve over $\chi^2$-test for sufficiently large, albeit finite, $\eta^2$s. However, it indeed is possible to do so for smaller $\eta^2$s. The biggest gain, not surprisingly, occurs when $\eta=0$.

Let $K$ be the number of connected components in $G$, and $G_k=(V_k,E_k)$, $k=1,\ldots, K$, the components so that $G=\cup_k G_k$. In this setting, $L(G)$ has exactly $K$ zero eigenvalues corresponding to eigenvectors
$$\bv_k:=|V_k|^{-1/2}(\II(v\in V_k))_{v\in V},\qquad k=1,\ldots, K,$$
where $\II$ is the characteristic function that takes value 1 if the the condition holds and zero otherwise. Thus,
$$
\Theta(0)=\left\{\bmu=c_1\bv_1+\cdots+c_K\bv_K: c_1,\ldots,c_K\in \RR\right\},
$$
is a $K$ dimensional linear subspace of $\RR^{|V|}$. To test against the effect $\bmu\neq 0$ then amounts to testing against $(c_1,\ldots,c_K)^\top\neq {\bf 0}$. By Neyman-Pearson Lemma, the likelihood ratio test is the most power for such a purpose. More specifically, it is not hard to derive the likelihood ratio test statistic
\begin{equation}
\label{eq:s1s2}
R:=\sum_{k=1}^K \left(\bX^\top\bv_k\right)^2=\sum_{k=1}^K \left[|V_k|^{-1}\left(\sum_{v\in V_k} X_v\right)^2\right].
\end{equation}
Under $H_0$, it is not hard to see that $R\sim \chi^2_K$, so that a $\alpha$ level test would reject $H_0$ if and only if $R\ge \chi^2_{K,\alpha}$. As before, we denote this test by $\varphi_{R,\alpha}$, or $\varphi_R$ for short. We note that when $G$ is connected, that is $K=1$, $\varphi_R$ is equivalent to the $z$-test $\varphi_z$. On the other hand, under $H_a$ with the overall effect $\bmu\in \Theta(0)$, we get $L\sim \chi^2_K(\|\bmu\|^2)$. Thus, $R$ is consistent for testing against $\bmu\in \Theta(0)$ if $\|\bmu\|^2\gg K^{1/2}$. It turns out this is not only the best we can do for $\Theta(0)$, but also for $\Theta(\eta^2)$ with a sufficiently small $\eta^2$, in general.

\begin{theorem}
\label{th:s1s2optimal}
Let $G=(V,E)$ be the union of $K$ connected and non-overlapped subgraphs, and $\eta_{\min}^2(G)$ be the smallest nonzero eigenvalue of its Laplacian $L(G)$. Then, for any $\eta^2=O(\eta_{\min}^2)$, $\varphi_R$ is consistent for testing against any $\bmu\in \Theta(\eta^2)$ such that $\|\bmu\|^2\gg K^{1/2}$ in that $\beta(\varphi_R;\bmu)\to 0$.
\end{theorem}

Recall that $\Theta(0)\subset \Theta(\eta_{\min}^2)$ and there is no consistent test against $\bmu\in \Theta(0)$ obeying $\|\bmu\|^2=O(K^{1/2})$. Thus Theorem \ref{th:s1s2optimal} shows the optimality of $\varphi_R$ for testing against $\bmu\in \Theta(\eta_{\min}^2)$. Comparing the required strength of $\bmu$ characterized by Theorems \ref{th:s2optimal} and \ref{th:s1s2optimal}, we can see the tremendous advantage of knowing that an effect $\bmu$ is sufficiently smooth with respect to $G$, e.g., $\bmu^\top L(G)\bmu\le \eta_{\min}^2$.

However, it is also clear from the above discussion that different tests may be needed to fully exploit the smoothness of $\bmu$. Although it is plausible that an overall effect is smooth with respect to $G$, such knowledge is rarely known beforehand. Naturally, one may ask if there is an agnostic approach that does not require such knowledge yet can still automatically exploit the potential smoothness of a putative effect, a task akin to adaption in nonparametric testing \citep[see, e.g.,][]{IngsterSuslina2003}. To this end, we consider a class of test statistics designed to account for different levels of smoothess:
\begin{equation}
\label{eq:deftlam}
T_\lambda:={\bX^\top (I+\lambda L(G))^{-1}\bX - \tr[(I+\lambda L(G))^{-1}]\over \left\{\tr[(I+\lambda L(G))^{-2}]\right\}^{1/2}},
\end{equation}
where $\lambda\ge 0$ is a regularization parameter. The test statistic $T_\lambda$ is a normalized version of the quadratic form:
$$
\bX^\top (I+\lambda L(G))^{-1}\bX,
$$
which can be viewed as the $\chi^2$ statistic regularized by the graph Laplacian $L(G)$. In particular, when $\lambda=0$, $T_0$ is a normalized version of the $\chi^2$ statistic and therefore is the most powerful for detecting effects that are not necessarily smooth with respect to $G$. On the other hand, when $\lambda\to \infty$, $T_\infty$ is a normalized version of the likelihood ratio statistic defined in (\ref{eq:s1s2}) which is most powerful for testing against a sufficiently smooth effect $\bmu$. In general, it is expected that different tuning parameters are suitable for testing against effects of different levels of smoothness.

Not knowing the exact smoothness of $\bmu$, we seek the maximum over the whole class of test statistics, leading to the following test statistic:
\begin{equation}
\label{eq:deftmax}
T_{\max}=\max_{\lambda\ge 0} T_\lambda.
\end{equation}
In general, the distribution of $T_{\max}$ under $H_0$ may not be computed analytically. However, it can be readily evaluated by Monte Carlo simulation, or through permutation test in the context of gene set enrichment analysis. Denote by $q_{\alpha}$ the $1-\alpha$ quantile of the null distribution of $T_{\max}$, and we proceed to reject $H_0$ if and only if $T_{\max}>q_{\alpha}$. As usual, we shall hereafter denote this test by $\varphi_{T,\alpha}$, or $\varphi_T$ for short, when no confusion occurs.

\section{Average-Case Analysis}
\label{sec:ave}

To appreciate the merits and understand the operating characteristics of the proposed test statistic $T_{\max}$, it is illuminating to begin with the case when $G$ is a random graph, more specifically, an Erd\"os-R\'{e}nyi graph. Under the Erd\"os-R\'enyi model $ER(n,p)$, first introduced in 1959 \citep{ErdosRenyi1959}, a random graph of $n$ nodes is constructed by connecting each pair of nodes randomly: each edge is included in the graph with probability $p$ independently. Assuming that the underlying graph $G$ follows an Erd\"os-R\'enyi model, we can work out an explicit form for the asymptotic distribution of $T_{\max}$. Denote by $\bar{\mu}=\bmu^\top \one/n$ the average of the coordinates of $\bmu$, and $\bmu_c=\bmu-\bar{\mu} \one$ the centered version of $\bmu$.

\begin{theorem}
\label{pr:ernull}
Let $G_n$ be a sequence of Erd\"os-R\'enyi graphs with $n$ nodes and a fixed probability of edge inclusion $p\in (0,1)$, and $T_{\max}$ be defined by (\ref{eq:deftlam}) and (\ref{eq:deftmax}). Assume that $\bX\sim N(\bmu,I)$ such that
$$
\delta_1^2=\lim_{n\to\infty}{1\over \sqrt{n-1}}\|\bmu_c\|^2,\qquad {\rm and}\qquad  \delta_2=\lim_{n\to\infty}\sqrt{n}\bar{\mu}.
$$
Then
\begin{equation}
\label{eq:knHa}
T_{\max}\to_d\left\{\begin{array}{ll}(2Y_1^2+(Y_2^2-1)^2)^{1/2}& {\rm if\ } Y_1>0, Y_2^2>1\\\max\left\{\sqrt{2}Y_1,Y_2^2-1\right\} & {\rm otherwise}\end{array}\right.,\qquad {\rm as}\quad n\to \infty,
\end{equation}
where $Y_1\sim N(\delta_1^2,1)$ and $Y_2\sim N(\delta_2,1)$ are two independent normal random variables. In particular, if $\bX\sim N(0,I)$, then
\begin{equation}
\label{eq:knH0}
T_{\max}\to_d\left\{\begin{array}{ll}(2Z_1^2+(Z_2^2-1)^2)^{1/2}& {\rm if\ } Z_1>0, Z_2^2>1\\\max\left\{\sqrt{2}Z_1,Z_2^2-1\right\} & {\rm otherwise}\end{array}\right.,\qquad {\rm as}\quad n\to \infty,
\end{equation}
where $Z_1$ and $Z_2$ are two independent standard normal random variables.
\end{theorem}

\begin{figure}[htbp]
\begin{center}
\includegraphics[width=1.1\textwidth]{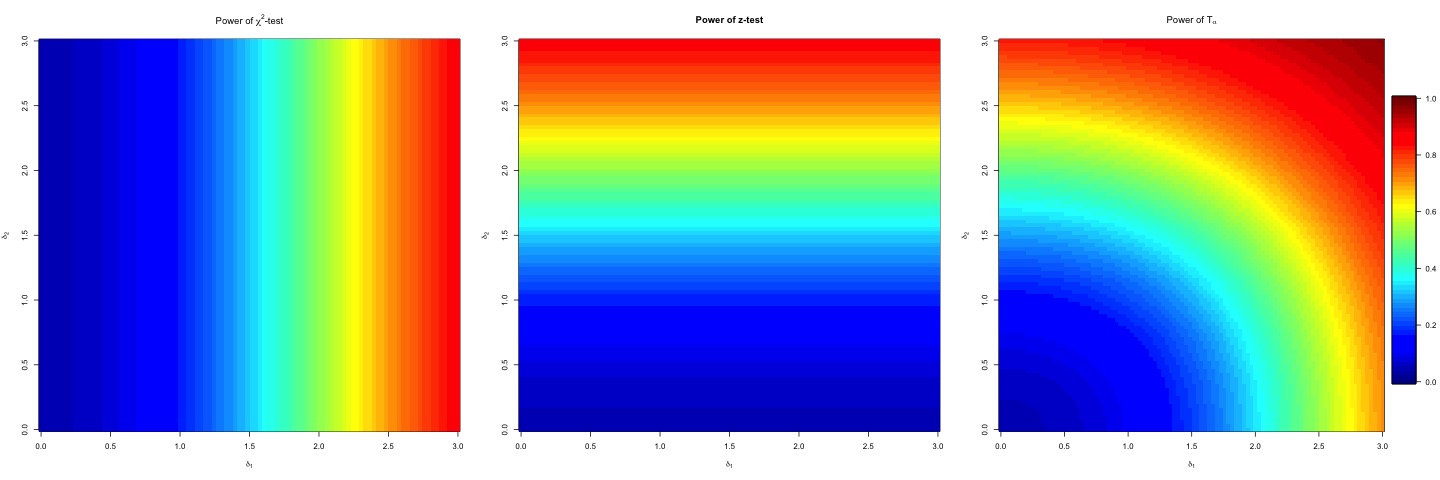}
\caption{Power of $z$-test, $\chi^2$-test and $T_{\max}$ based test $\varphi$, as functions of $\delta_1$ and $\delta_2$, for complete graphs.}
\label{fig:power-kn}
\end{center}
\end{figure}

Equation (\ref{eq:knH0}) allows us to compute more explicitly the critical value of a test based on $T_{\max}$ at a prescribed significance level, at least in an asymptotic sense. Together with (\ref{eq:knHa}), this allows us to more precisely characterize the (asymptotic) power of $\varphi_T$. In particular, the power of the $5\%$-level test, as a function of $\delta_1$ and $\delta_2$, is shown in the rightmost panel of Figure \ref{fig:power-kn}. It is also instructive to compare the power of the test with that of the $\chi^2$-test and $z$-test. As mentioned before, the $\chi^2$-test is known to be minimax optimal for testing against all possible effect $\bmu\neq {\bf 0}$ whereas $z$-test is the most powerful for testing against a constant effect of the form $\bmu\propto{\bf 1}$. The power of $\chi^2$ and $z$ tests at $5\%$ level, again as functions of $\delta_1$ and $\delta_2$, is also given in Figure \ref{fig:power-kn} for comparison.

\begin{figure}
    \centering
    \begin{subfigure}[b]{0.48\textwidth}
    \centering
        \includegraphics[width=\textwidth]{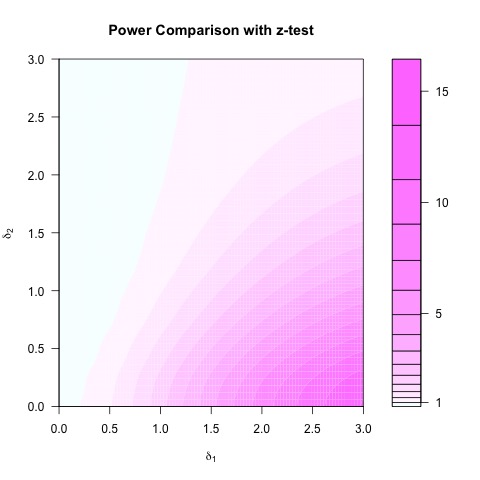}
    \end{subfigure}
    \begin{subfigure}[b]{0.48\textwidth}
    \centering
        \includegraphics[width=\textwidth]{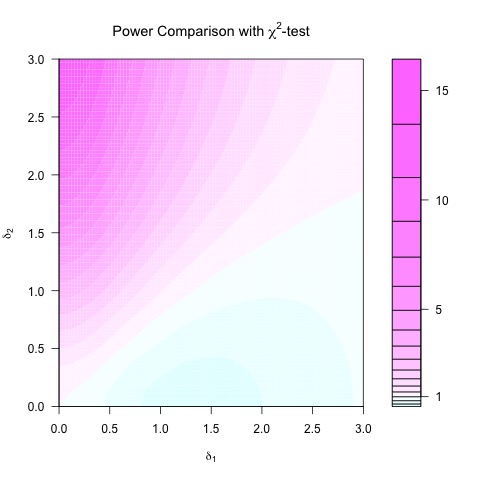}
    \end{subfigure}
    \caption{Relative power of $\varphi_T$ with respect to the $z$-test (left panel) and $\chi^2$-test (right panel). In the pink region of each panel, $\varphi_T$ outperforms the alternative test.}
    \label{fig:power-comp-kn}
\end{figure}

For further comparison, we plot in Figure \ref{fig:power-comp-kn} the ratio of the power of $\varphi_T$ over that of the $\chi^2$ and $z$ tests, again as functions of $\delta_1$ and $\delta_2$. The minimum ratios are $85.7\%$ and $61.2\%$ respectively indicating that $\varphi_T$ is at least $85.7\%$ as powerful as the $z$-test, and $61.2\%$ as powerful as the $\chi^2$-test. On the other hand, the maximum of both ratios can be arbitrarily large suggesting $\varphi_T$ can be arbitrarily more powerful than both the $\chi^2$ and $z$ tests. Therefore, in absence of further information about the putative effect $\bmu$, $\varphi_T$ could be more preferable to either $\chi^2$ or $z$ test.

In fact, not only superior to $\chi^2$ and $z$ tests, $\varphi_T$ can also be shown, in a certain sense, to be optimal. More specifically,

\begin{theorem}
\label{th:er}
Let $G_n$ be a sequence of Erd\"os-R\'enyi graphs with $n$ nodes and a fixed probability of edge inclusion $p\in (0,1)$. For any $\eta^2\ge 0$, $\varphi_T$ is consistent in testing against $\bmu\in \Theta_{G_n}(\eta^2)$ such that 
$$
\|\bmu\|^2\gg r^2_{\rm ER}(\eta^2):= \left\{\begin{array}{ll}n^{1/2}& {\rm if\ }\eta\ge n^{3/4}\\ \eta^{2}/n& {\rm if\ } n^{1/2}\le \eta\le n^{3/4}\\ 1& {\rm if\ }\eta\le n^{1/2}\end{array}\right.,
$$
in that the Type II error $\beta(\varphi_T;\bmu)\to 0$. On the other hand, there exists a constant $c>0$ such that for any $\eta^2\ge 0$, and any $\alpha$-level ($0<\alpha<1$) test $\Psi$ based on data $(X_v)_{v\in V}$,
$$
\liminf_{|V|\to\infty}\esssup_{\bmu\in \Theta_{G_n}(\eta^2): \|\bmu\|^2\ge cr_{\rm ER}^2(\eta^2)}\beta(\Psi;\bmu)>0.
$$
\end{theorem}

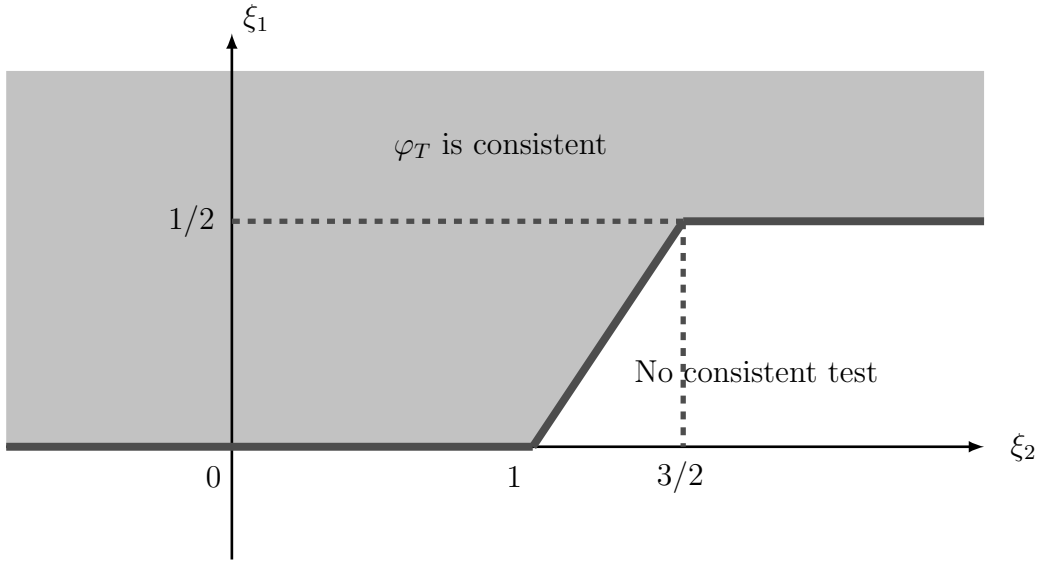
\begin{figure}[htbp]
\begin{center}
\begin{tikzpicture}
 \begin{scope}
       \fill[white,fill opacity=0.8] (-3,-1) rectangle (10,6); 
       \fill[black!30!white,fill opacity=0.8] (-3,0) -- (4,0) -- (6,3) -- (10,3) -- (10,5) -- (-3,5) --cycle ; 
       \path[-latex,draw,line width=1pt] (-3,0) -- (10,0);
       \path[-latex,draw,line width=1pt] (0,-1.5) -- (0,5.5);
       
       \path[draw,line width=3pt,black!70!white] (-3,0) -- (4,0);
       \path[draw,line width=3pt,black!70!white] (4,0) -- (6,3);
       \path[draw,line width=3pt,black!70!white] (6,3) -- (10,3);
       
       \path[draw,line width=2pt,black!70!white,dashed] (6,3) -- (6,0);
       \path[draw,line width=2pt,black!70!white,dashed] (0,3) -- (6,3);
 \end{scope}
 \draw[thick,black](-0.5,-0.4)node[right]{$0$};
  \draw[thick,black](3.5,-0.4)node[right]{$1$};
  \draw[thick,black](5.5,-0.4)node[right]{$3/2$};
  \draw[thick,black](-1,3)node[right]{${1/2}$};
  
  \draw[thick,black](10.2,0)node[right]{$\xi_2$};
  \draw[thick,black](0,5.7)node[right]{$\xi_1$};

  \draw[thick,black](2,4)node[right]{$\varphi_T$ is consistent};
  \draw[thick,black](5.2,1)node[right]{No consistent test};
\end{tikzpicture}
\end{center}
\caption{Detection Boundary for Erd\"os-R\'enyi Graphs}
\label{fig:er}
\end{figure}

Theorem \ref{th:er} shows that, if $\eta^2$ is known in advance, then there is no consistent test for effect $\bmu\in \Theta(\eta^2)$ such that $\|\bmu\|^2=O(r^2_{\rm ER}(\eta^2))$; and conversely, if $\|\bmu\|^2\gg r^2_{\rm ER}(\eta^2)$, then $\varphi_T$ is consistent. Putting it differently, the test $\varphi_T$ attains the optimal {\it detection boundary} $r_{\rm ER}^2(\eta^2)$ for any effects for a given level ($\eta^2$) of smoothness although it does not assume such knowledge. It is instructive to consider the case when $\|\bmu\|^2=n^{\xi_1}$ and $\bmu^\top L(G_n)\bmu=n^{\xi_2}$. Theorem \ref{th:er} shows that the boundary for $\bmu$ to be consistently testable can be given by the diagram in Figure \ref{fig:er}.

One can think of Erd\"os-R\'enyi model as a way to assign probability over all graphs with $n$ nodes. Theorem \ref{th:er} shows that the set of graphs for which the test $\varphi_T$ can achieve the optimal detection boundary $r_{\rm ER}^2(\eta^2)$ has probability tending to one under this measure. In other words, $\varphi_T$ is minimax optimal for almost all graphs. The detection boundary $r_{\rm ER}^2(\eta^2)$ also characterizes the extent to which $\varphi_T$ indeed can  provide improved performance depending the potential smoothness of an effect without assuming such knowledge is available to us. Conceptually, our treatment of Erd\"os-R\'enyi model is akin to an average-case analysis. On the other hand, it may also be of interest to investigate the performance of $T_{\max}$ for specific graphs, which we shall do in the next section.

\section{General Performance Bounds}
\label{sec:spe}

To complement our treatment to random graphs, we now investigate the performance of $T_{\max}$ for a specific graph $G=(V,E)$, again with the focus on the case when $|V|$ is large. Precise characterization of the operating characteristics of $T_{\max}$ becomes elusive for general graphs because closed form expressions of its asymptotic distributions such as those presented in Theorem \ref{th:er} are no longer available. Nonetheless, we shall derive in this section generally applicable performance bounds for $\varphi_T$.

More specifically, consider the following equation in variable $x\ge 0$:
\begin{equation}
\label{eq:master}
x^2=(\log\log|V|)\cdot\tr\left[I+{x\over 2\eta^2}L(G)\right]^{-2}.
\end{equation}
It is clear that as $x$ increases from zero to infinity, so does the left hand side of (\ref{eq:master}); while the right hand side decreases from $|V|\log\log |V|$ to zero, so that the equation has a unique solution between $0$ and $\sqrt{|V|\log\log |V|}$, hereafter denoted by $x_\ast(G,\eta^2)$. The following theorem shows that $\varphi_T$ is consistent in testing against any $\bmu\in \Theta_G(\eta^2)$ such that $\|\bmu\|^2\gg x_\ast(G, \eta^2)$.

\begin{theorem}
\label{th:power}
Let $\lambda_{\min}$ be the smallest nonzero eigenvalue of the Laplacian matrix $L(G)$ of $G=(V, E)$. Assume that $\log(1/\lambda_{\min})=O(\log |V|)$. Then $\beta(\varphi_T;\bmu)\to 0$ for any $\bmu\in \RR^{|V|}$ such that
\begin{equation}
\label{eq:altercond}
\sup_{\alpha\ge 0}{\bmu^\top (I+\alpha L)^{-1}\bmu\over \left\{\tr[(I+\alpha L)^{-2}]\right\}^{1/2}}\gg \log\log |V|.
\end{equation}
In particular, if $\bmu\in \Theta_G(\eta^2)$, then $\beta(\varphi_T;\bmu)\to 0$ whenever
\begin{equation}
\label{eq:bdgeneral}
\|\bmu\|^2\gg x_\ast(G,\eta^2),
\end{equation}
where $x_\ast(G,\eta^2)$ is the solution to (\ref{eq:master}).
\end{theorem}

Several observations follow immediately from Theorem \ref{th:power}. Recall that
$$x_\ast(G,\eta^2)\le \sqrt{|V|\log\log |V|},$$
so that $\varphi_T$ is consistent for testing against any $\bmu\in \RR^{|V|}$ such that 
\begin{equation}
\label{eq:bdlarge}
\|\bmu\|^2\gg \sqrt{|V|\log\log|V|},
\end{equation}
in the light of (\ref{eq:bdgeneral}).  On the other hand, by fixing $\alpha=+\infty$ on the right hand side of (\ref{eq:altercond}) we get $\beta(\varphi_T;\bmu)\to 0$ for any $\bmu\in \Theta_G(0)$ such that
\begin{equation}
\label{eq:bdsmall}
\|\bmu\|^2\gg \sqrt{K\log\log |V|},
\end{equation}
where $K$ is the number of non-overlap connected components in $G$. In fact, using the same argument as that for Theorem \ref{th:s1s2optimal}, we can show that $\varphi_T$ is also consistent in testing against any $\bmu\in \Theta_G(\lambda_{\min})$ such that (\ref{eq:bdsmall}) holds.

The performance bounds (\ref{eq:bdlarge}) and (\ref{eq:bdsmall}) are nearly optimal in that they differ from the optimal bounds given by Theorems \ref{th:s2optimal} and \ref{th:s1s2optimal} only by an iterated logarithmic factor in $|V|$. Such an iterated logarithmic factor also exists for general $\eta^2$s, as a result of the presence of the $\log\log |V|$ term on the right hand side of (\ref{eq:master}) or (\ref{eq:altercond}). In the light of the average-case analysis presented in the previous section, we know that such an extra iterated logarithmic factor is unnecessary for almost all graphs under Erd\"os-R\'enyi model. However, as we shall we see below that for certain type of graphs, this extra factor is indeed necessary, and hence unavoidable here because of the generality of our results.

We now consider several fundamental types of graphs to demonstrate that these general performance bounds are indeed (nearly) optimal.

\begin{figure}[htbp]
\begin{minipage}[c]{0.49\textwidth}
\begin{center}
\begin{tikzpicture}
 \begin{scope}
       \fill[white,fill opacity=0.8] (-3,-3) rectangle (3,3); 
       
       \path[draw,line width=2pt,blue!50!white] ({1.5*sqrt(3)},-1.5) -- (-{1.5*sqrt(3)},1.5);
       
       \path[draw,line width=2pt,blue!50!white] ({1.5*sqrt(3)},1.5) -- (-{1.5*sqrt(3)},-1.5);
       
       \path[draw,line width=2pt,blue!50!white] (0,-3) -- (0,3);
       
       \draw[fill=black!70!white] (0,0) ellipse (0.1 and 0.1);
       \draw[fill=black!70!white] (0,3) ellipse (0.1 and 0.1);
       \draw[fill=black!70!white,rotate=180] (0,3) ellipse (0.1 and 0.1);
       \draw[fill=black!70!white,rotate=60] (0,3) ellipse (0.1 and 0.1);
       \draw[fill=black!70!white,rotate=120] (0,3) ellipse (0.1 and 0.1); 
       \draw[fill=black!70!white,rotate=-120] (0,3) ellipse (0.1 and 0.1); 
       \draw[fill=black!70!white,rotate=-60] (0,3) ellipse (0.1 and 0.1);   
 \end{scope}
 \draw[thick,black](0.3,0)node[right]{$V_1$};
 \draw[thick,black](-0.3,3.5)node[right]{$V_2$};
 \draw[thick,black](-3.4,1.6)node[right]{$V_3$};
 \draw[thick,black](-3.4,-1.6)node[right]{$V_4$};
 \draw[thick,black](-0.3,-3.5)node[right]{$V_5$};
 \draw[thick,black](2.75,-1.6)node[right]{$V_6$};
 \draw[thick,black](2.75,1.6)node[right]{$V_7$};
 \end{tikzpicture}
\end{center}
\caption{Star Graph}
\label{fig:star}
\end{minipage}
\begin{minipage}[c]{0.49\textwidth}
$$
L(S_n)=\left[\begin{array}{cccccc}n-1&-1&-1&\ldots&\ldots&-1\\-1&1&0&0&\ldots&0\\-1&0&1&0&\ldots&0\\ \vdots&\vdots&\ddots&\ddots&\ddots&\vdots\\-1&\ldots&\ldots&0&1&0\\-1&\ldots&\ldots&\ldots&0&1\end{array}\right].
$$
\end{minipage}
\end{figure}
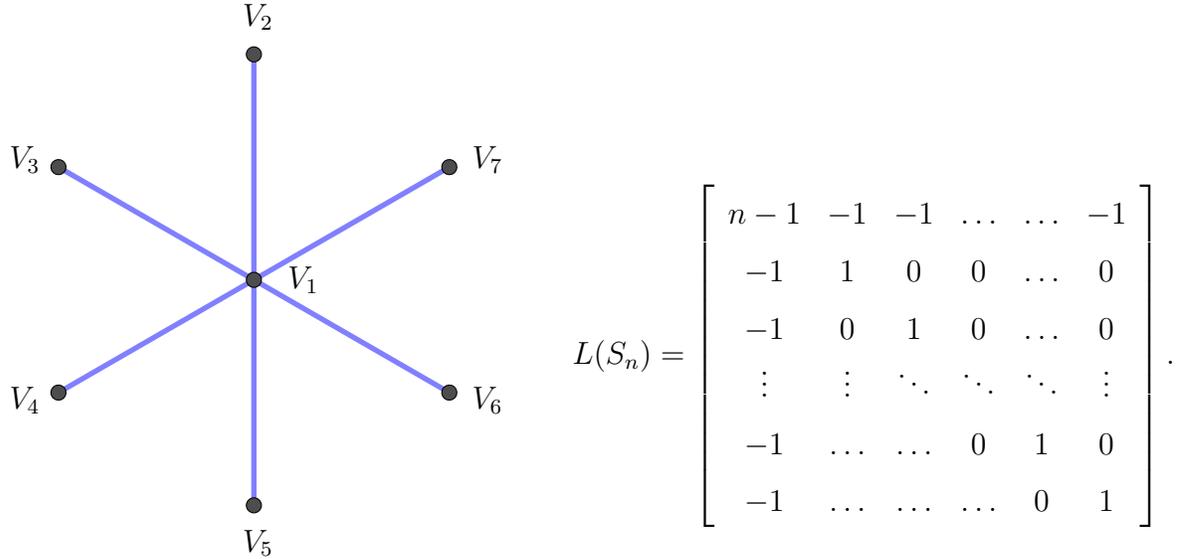

\paragraph{Star Graph.} Our first example is the so-called star graph where one node is connected with all the remaining nodes, as show in Figure \ref{fig:star}. The Laplacian matrix of a star graph with $n$ vertices, denoted by $S_n$, can also be given explicitly.

It is well known that in this case, the eigenvalues of the Laplacian are
$$
n=\lambda_1(L(S_n))>\lambda_2(L(S_n))=\cdots=\lambda_{n-1}(L(S_n))=1,\qquad {\rm and}\qquad \lambda_n(L(S_n))=0.
$$
It is not hard to derive from (\ref{eq:altercond}) that $\varphi_T$ is consistent for testing against any $\bmu\in \Theta_{S_n}(\eta^2)$ such that
$$
\|\bmu\|^2\gg \left\{\begin{array}{ll}(n\log\log n)^{1/2}& {\rm if\ }\eta\ge n^{1/4}\\ \eta^2(\log\log n)^{1/2}& {\rm if\ } 1\le \eta\le n^{1/4}\\ (\log\log n)^{1/2}& {\rm if\ }\eta\le 1\end{array}\right..
$$
This bound turns out to be optimal up to the iterated logarithmic factor.

\begin{theorem}
\label{th:star}
For a star graph $S_n$, $\beta(\varphi_T;\bmu)\to 0$ for any $\bmu\in \Theta_{S_n}(\eta^2)$ such that
$$
\|\bmu\|^2\gg r_{S_n}^2(\eta^2):= \left\{\begin{array}{ll}n^{1/2}& {\rm if\ }\eta\ge n^{1/4}\\ \eta^2& {\rm if\ } 1\le \eta\le n^{1/4}\\ 1& {\rm if\ }\eta\le 1\end{array}\right..
$$
Conversely, there exists a constant $c>0$ such that for any $\eta^2\ge 0$, and any $\alpha$-level ($0<\alpha<1$) test $\Psi$ based on data $(X_v)_{v\in V}$,
$$
\liminf_{|V|\to\infty}\esssup_{\bmu\in \Theta_{S_n}(\eta^2): \|\bmu\|^2\ge cr_{S_n}^2(\eta^2)}\beta(\Psi;\bmu)>0.
$$
\end{theorem}

\begin{figure}[htbp]
\begin{minipage}[c]{0.49\textwidth}
\begin{center}
\begin{tikzpicture}
 \begin{scope}
       \fill[white,fill opacity=0.8] (-3,-3) rectangle (3,3); 
       \path[draw,line width=2pt,blue!50!white] ({1.5*sqrt(3)},-1.5) -- ({1.5*sqrt(3)},1.5);
       \path[draw,line width=2pt,blue!50!white] (-{1.5*sqrt(3)},-1.5) -- (-{1.5*sqrt(3)},1.5);
       \path[draw,line width=2pt,blue!50!white] ({1.5*sqrt(3)},-1.5) -- (0,-3);
       \path[draw,line width=2pt,blue!50!white] (-{1.5*sqrt(3)},-1.5) -- (0,-3);
       \path[draw,line width=2pt,blue!50!white] (-{1.5*sqrt(3)},1.5) -- (0,3);
       \path[draw,line width=2pt,blue!50!white] ({1.5*sqrt(3)},1.5) -- (0,3);
       \draw[fill=black!70!white] (0,3) ellipse (0.1 and 0.1);
       \draw[fill=black!70!white,rotate=180] (0,3) ellipse (0.1 and 0.1);
       \draw[fill=black!70!white,rotate=60] (0,3) ellipse (0.1 and 0.1);
       \draw[fill=black!70!white,rotate=120] (0,3) ellipse (0.1 and 0.1); 
       \draw[fill=black!70!white,rotate=-120] (0,3) ellipse (0.1 and 0.1); 
       \draw[fill=black!70!white,rotate=-60] (0,3) ellipse (0.1 and 0.1);   
 \end{scope}
 \draw[thick,black](-0.3,3.5)node[right]{$V_1$};
 \draw[thick,black](-3.4,1.6)node[right]{$V_2$};
 \draw[thick,black](-3.4,-1.6)node[right]{$V_3$};
 \draw[thick,black](-0.3,-3.5)node[right]{$V_4$};
 \draw[thick,black](2.75,-1.6)node[right]{$V_5$};
 \draw[thick,black](2.75,1.6)node[right]{$V_6$};
 \end{tikzpicture}
\end{center}
\caption{Cycle Graph $C_6$}
\label{fig:cycle}
\end{minipage}
\begin{minipage}[c]{0.49\textwidth}
$$
L(C_n)=\left[\begin{array}{cccccc}2&-1&0&\ldots&\ldots&-1\\-1&2&-1&0&\ldots&0\\0&-1&2&-1&\ldots&0\\ \vdots&\vdots&\ddots&\ddots&\ddots&\vdots\\0&\ldots&\ldots&-1&2&-1\\-1&\ldots&\ldots&\ldots&-1&2\end{array}\right]
$$
\end{minipage}
\end{figure}
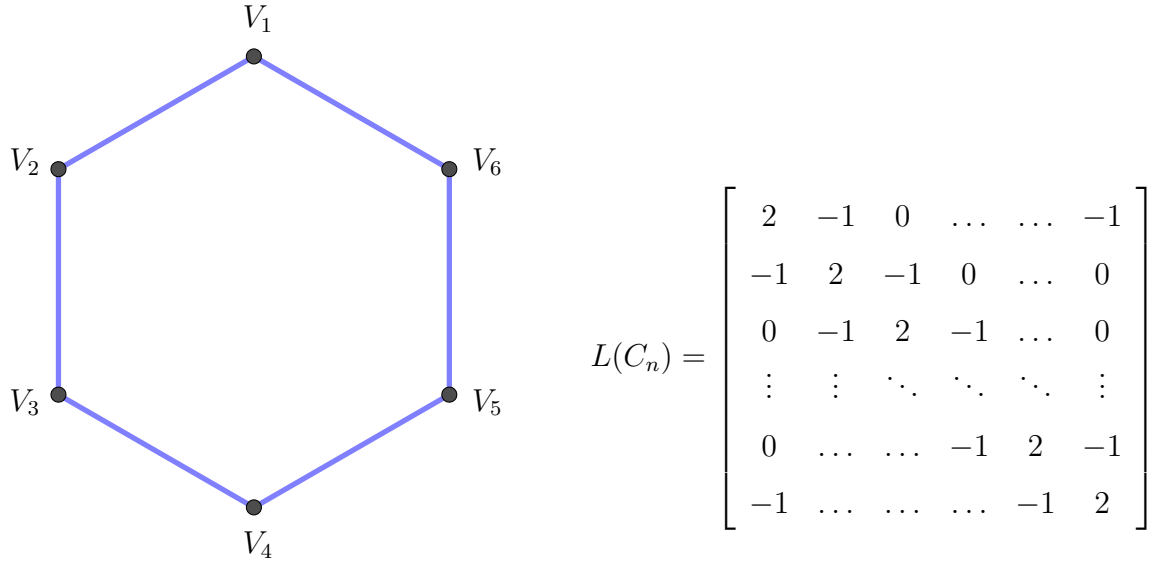

\paragraph{Cycle Graphs.} We now consider another example to show that at least for some types of graphs, the extra iterated logarithmic factor cannot be removed. In the so-called cycle graphs, the nodes form a ring and each node is connected with its two neighbors, as shown in Figure \ref{fig:cycle}. A cycle graph with $n$ vertices is commonly denoted by $C_n$. Its Laplacian $L(C_n)$ can be given explicitly.

It is well known that the eigenvalues of $L(C_n)$ is given by
$$
2-2\cos\left({2\pi k\over n}\right)=4\sin^2\left({\pi k\over n}\right),\qquad k=1,\ldots,n.
$$
See, e.g, \cite{BrouwerHaemers2012}. Thus,
$$
\tr[(I+x L(C_n)/2\eta^2)^{-2}]=\sum_{k=1}^{n}\left[1+{2x\over \eta^2}\sin^2\left({\pi k\over n}\right)\right]^{-2}\asymp \sum_{k=1}^{n}\left[1+{x\over \eta^2}\left({k\over n}\right)^2\right]^{-2}.
$$
Hereafter $a_n\asymp b_n$ means $a_n/b_n$ is bounded away from $0$ and $+\infty$. Let $k_\ast(x)={n\eta x^{-1/2}}$. If $1\le k_\ast(x)\le n$, then
$$
\sum_{k=1}^{n}\left[1+{x\over \eta^2}\left({k\over n}\right)^2\right]^{-2}\asymp k_\ast(x),
$$
which immediately implies that
$$
x_\ast(C_n;\eta^2)\asymp (n\eta\log\log n)^{2/5},
$$
provided that
$$1\le k_\ast(x_\ast(C_n;\eta^2))\le n.$$
By Theorem \ref{th:power}, we get

\begin{corollary}
For any $\eta^2\ge 0$ and $\bmu\in \Theta_{C_n}(\eta^2)$, $\varphi_T$ is consistent in that $\beta(\varphi_T;\bmu)\to 0$ if
\begin{equation}
\label{eq:cycle}
\|\bmu\|^2\gg \left\{\begin{array}{ll}(n\log\log n)^{1/2}& {\rm if\ }\eta\ge (n\log\log n)^{1/4}\\ (n\eta\log\log n)^{2/5}& {\rm if\ }n^{-1}(\log\log n)^{1/4}\le\eta<(n\log\log n)^{1/4}\\(\log\log n)^{1/2} &{\rm if\ }\eta<n^{-1}(\log\log n)^{1/4}\end{array}\right..
\end{equation}
\end{corollary}

It turns out that this performance bound is, in a certain sense, optimal.

\begin{theorem}
\label{th:cycle}
There exists a constant $c>0$ such that for any $-1< a<b<1/4$, and any $\alpha$-level ($0<\alpha<1$) test $\Psi$,
$$
\liminf_{n\to\infty}\sup_{\substack{\bmu\in \Theta(\eta^2):\|\bmu\|^2\ge cx_\ast(C_n;\eta^2)\\ \log\eta/\log n\in (a,b)}}\beta(\Psi,\bmu)>0.
$$
\end{theorem}

\begin{figure}[htbp]
\begin{center}
\begin{tikzpicture}
 \begin{scope}
       \fill[white,fill opacity=0.8] (-3,-1) rectangle (10,6); 
       \fill[black!30!white,fill opacity=0.8] (-3,0) -- (-2,0) -- (9,3) -- (10,3) -- (10,5) -- (-3,5) --cycle ; 
       \path[-latex,draw,line width=1pt] (-3,0) -- (10,0);
       \path[-latex,draw,line width=1pt] (6,-1.5) -- (6,5.5);
       
       \path[draw,line width=3pt,black!70!white] (-3,0) -- (-2,0);
       \path[draw,line width=3pt,black!70!white] (-2,0) -- (9,3);
       \path[draw,line width=3pt,black!70!white] (9,3) -- (10,3);
       
       \path[draw,line width=2pt,black!70!white,dashed] (9,3) -- (9,0);
       \path[draw,line width=2pt,black!70!white,dashed] (6,3) -- (9,3);
 \end{scope}
 \draw[thick,black](-2.4,-0.4)node[right]{$-2$};
  \draw[thick,black](8.8,-0.4)node[right]{$1/2$};
  \draw[thick,black](5,3)node[right]{${1/2}$};
  
  \draw[thick,black](10.2,0)node[right]{$\xi_2$};
  \draw[thick,black](6,5.7)node[right]{$\xi_1$};
  
  \draw[thick,black](2,4)node[right]{$\varphi_T$ is consistent};
  \draw[thick,black](5.2,1)node[right]{No consistent test};
\end{tikzpicture}
\end{center}
\caption{Detection Boundary of Cycle Graph}
\label{fig:cyclebd}
\end{figure}
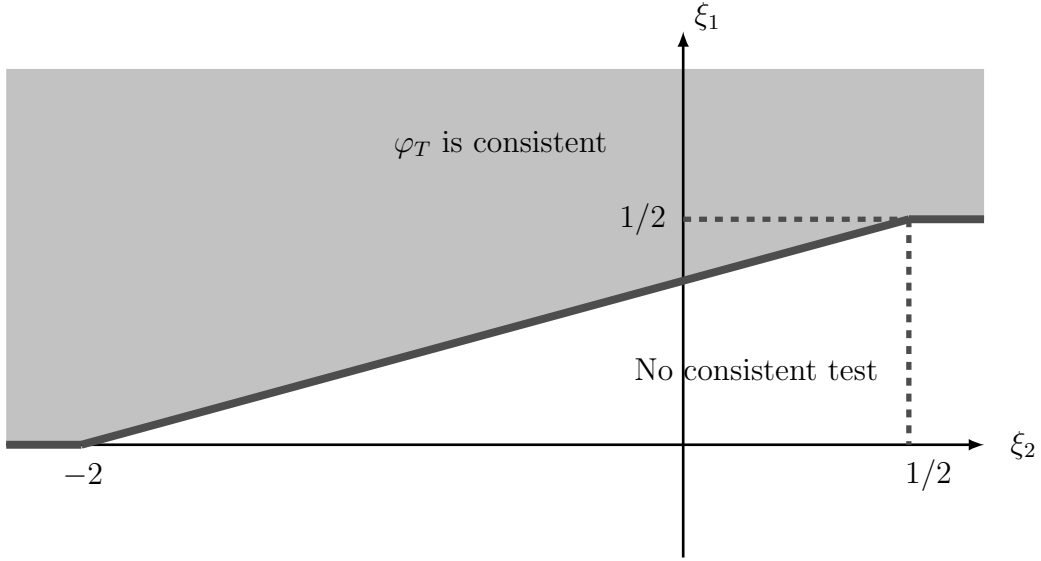

In other words, even if we know the smoothness index of $\bmu$ is between $n^{a}$ and $n^{b}$ for some $-1<a<b<1/4$, the best detection boundary can still be characterized by $x_\ast(C_n;\eta^2)$. As before, it is instructive to consider the case when $\|\bmu\|^2=n^{\xi_1}$ and $\bmu^\top L(C_n)\bmu=n^{\xi_2}$. The detection boundary in the $(\xi_1,\xi_2)$ plane for this case is shown in Figure \ref{fig:cyclebd}.

\paragraph{Lattice Graphs.} Our last example is the lattice graph. Consider a $d$-dimensional square lattice with size $m\times m\times\cdots\times m$ where at each lattice point, namely a point with integer coordinates $(i_1,i_2,\ldots,i_d)$ where $1\le i_k\le m$ ($k=1,2,\ldots, d$), a node is placed. Each node $(i_1,i_2,\ldots,i_d)$ is connected to its immediate neighbors $(i_1\pm 1, i_2,\ldots,i_d)$, $(i_1,i_2\pm 1,\ldots, i_d)$, $\ldots (i_1,i_2,\ldots,i_d\pm 1)$ if they are on the lattice.

Note that the lattice graph, denoted by $T_{m,d}$ can be viewed the Cartesian product $P_{m}\times \cdots\times P_{m}$ where $P_m$ is a path graph with $m$ nodes. Using the general result by \cite{fiedler73} for Cartesian product, we can write the eigenvalues of the Laplacian $L(T_{m,d})$ as
$$
\lambda_{j_1,\ldots,j_d}(L(T_{m,d}))=\lambda_{j_1}(L(P_{m}))+\cdots+\lambda_{j_d}(L(P_{m})),\qquad 0\le j_1,\ldots,j_d< m,
$$
where $\lambda_j(P_m)$ is the $j$th eigenvalue of $L(P_m)$. It is well known that
$$
\lambda_j(P_m)=4\sin^2\left({\pi j\over 2m}\right), \qquad j=0,\ldots,m-1.
$$
See, e.g, \cite{BrouwerHaemers2012}. Therefore,
$$
\lambda_{j_1,\ldots,j_d}=4\sin^2\left({\pi j_1\over 2m}\right)+\cdots+4\sin^2\left({\pi j_d\over 2m}\right).
$$
Following a similar argument as before, we can derive from Theorem \ref{th:power} that

\begin{corollary}
\label{co:lattice}
For any $\eta^2\ge 0$ and $\bmu\in \Theta_{T_{m,d}}(\eta^2)$, $\varphi_T$ is consistent in that $\beta(\varphi_T;\bmu)\to 0$ if
\begin{equation}
\label{eq:lattice}
\|\bmu\|^2\gg \left\{\begin{array}{ll}(n\log\log n)^{1/2}& {\rm if\ }\eta\ge (n\log\log n)^{1/4}\\ \eta^{2d\over 4+d}(n\log\log n)^{2\over 4+d}& {\rm if\ }n^{-1/d}(\log\log n)^{1/4}\le\eta<(n\log\log n)^{1/4}\\(\log\log n)^{1/2} &{\rm if\ }\eta<n^{-1/d}(\log\log n)^{1/4}\end{array}\right.,
\end{equation}
where $n=m^d$.
\end{corollary}

By the same argument as that for Theorem \ref{th:cycle}, it can also be shown that the detection rate given by (\ref{eq:lattice}) is indeed optimal and cannot be further improved. We omit the details here for brevity.

\section{Numerical Experiments}
\label{sec:sim}

We now present some numerical experiments to illustrate the merits of the proposed methodology and verify the theoretical findings reported earlier. In computing $T_{\max}$, we optimize over $\lambda$ using the function \texttt{nlm} in \texttt{R}, which is based on Newton method.

\subsection{Detection boundary}
We first conduct a set of simulation studies to verify the detection boundaries established by our theoretical analysis. To fix ideas, we set the the critical value to be the upper 5\% quantile of null distribution based on 1000 Monte Carlo simulations, which ensures that corresponding test is the 5\%-level test, up to Monte-Carlo error.

To demonstrate the adaptivity of the proposed test, we consider different combinations of values for the strength $\|\bmu\|^2$ and smoothness $\eta^2$. For a graph $G$, we simulated the effect $\bmu$ at a given $\bmu^\top L(G)\bmu$ and $\|\bmu\|^2$ as follows. Let $\bw_1,\ldots,\bw_n$ be the eigenvector of Laplacian matrix $L(G)$ corresponding to eigenvalues $\lambda_1\ge \ldots\ge \lambda_n=0$. We generated $\bmu$ of the following form:
$$
\bmu=\sum_{k=1}^n e_ku_k\bw_k
$$
where $e_k$s are Rademacher variables, i.e., $\PP(e_k=\pm 1)=1/2$, and
$$
u_k^2=\zeta_1\max(1-\zeta_2\lambda_k,0),
$$
where $\zeta_1$ and $\zeta_2$ are chosen such that
$$
\sum_{k=1}^n u_k^2=n^{\xi_1},\qquad{\rm and}\qquad \sum_{k=1}^n \lambda_ku_k^2=n^{\xi_2},
$$
for given values of $\xi_1$ and $\xi_2$.

\begin{figure}[htbp]
\begin{center}
\includegraphics[width=\textwidth]{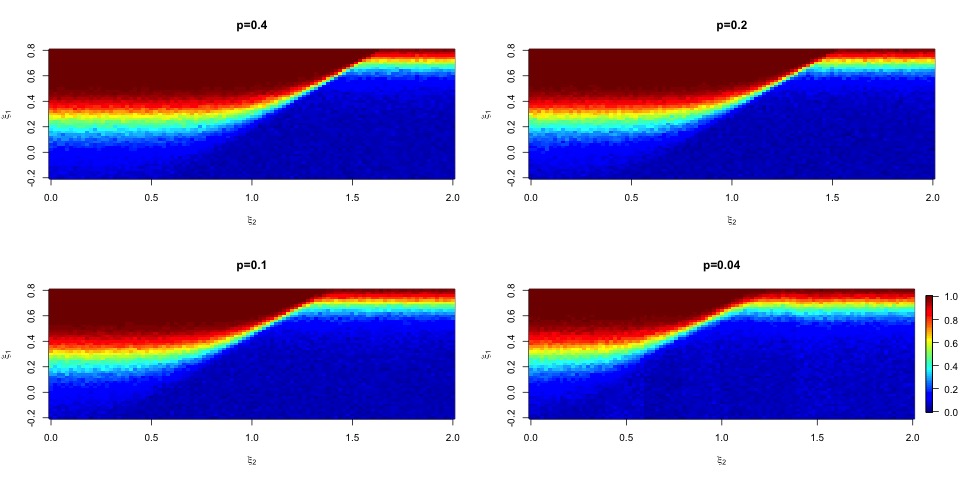}
\caption{Empirical detection boundary for random graphs following Erd\"os-R\'enyi model with 500 nodes and probability of edge inclusion at 0.4, 0.2, 0.1 and 0.01 respectively.}
\label{fig:randomsim}
\end{center}
\end{figure}

To assess the power of our method, we first consider Erd\"os-R\'enyi graphs with $n=500$ nodes and probability $p=0.4, 0.2, 0.1$ and $0.04$. For each value of $p$, $\xi_1$ and $\xi_2$, we simulated 500 graphs, and for each graph, we simulated $\bmu$ such that $\|\bmu\|^2=n^{\xi_1}$ and $\bmu^\top L(G)\bmu=n^{\xi_2}$ as described above. The observations were then generated and the frequency that the null hypothesis is rejected over these 500 graphs is given in Figure \ref{fig:randomsim}. Each plot in Figure \ref{fig:randomsim} was produced by repeating this experiment for combinations of 50 equally-spaced $\xi_1$ between 0 and 2, and $\xi_2$ between -0.2 and 0.8. It is clear from Figure \ref{fig:randomsim} that there is indeed a detection boundary which characterizes when an overall effect can be consistently tested, as predicted by our theoretical analysis. Furthermore, the empirical detection boundary agrees well with our theoretical results.

\begin{figure}[htbp]
\begin{center}
\includegraphics[width=\textwidth]{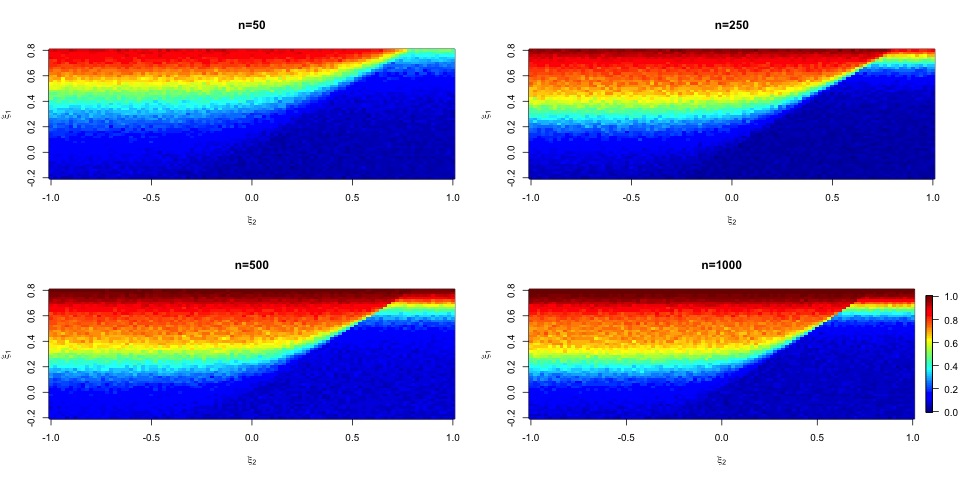}
\caption{Empirical detection boundary for star graph with 50, 250, 500 and 1000 nodes.}
\label{fig:starsim}
\end{center}
\end{figure}

\begin{figure}[htbp]
\begin{center}
\includegraphics[width=\textwidth]{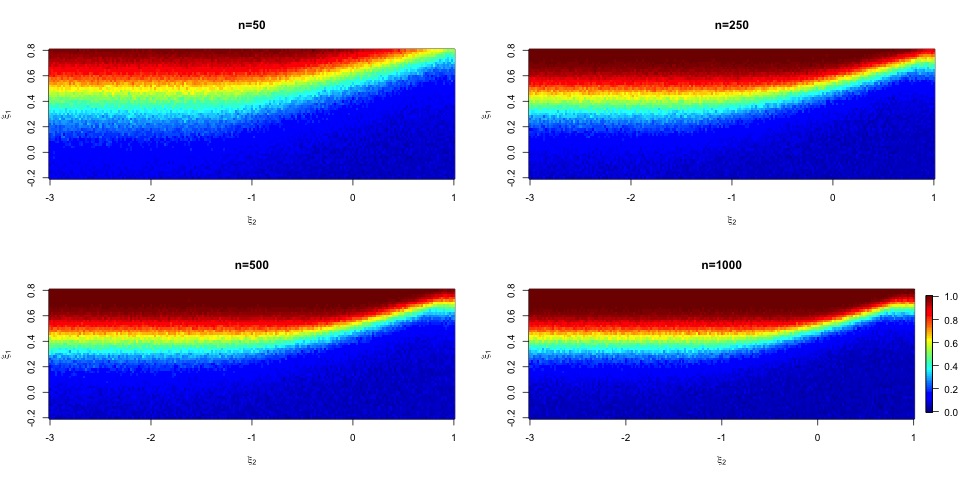}
\caption{Empirical detection boundary for cycle graph with 50, 250, 500 and 1000 nodes.}
\label{fig:cyclesim}
\end{center}
\end{figure}

We also conducted similar experiments for the star graph and cycle graph, each with $n=50, 250, 500$ or $1000$ nodes. The result, as shown in Figures \ref{fig:starsim} and \ref{fig:cyclesim}, again agrees well with our theoretical findings.

\subsection{Comparison with other test statistics}
\begin{figure}[htbp]
\begin{center}
\includegraphics[width=\textwidth]{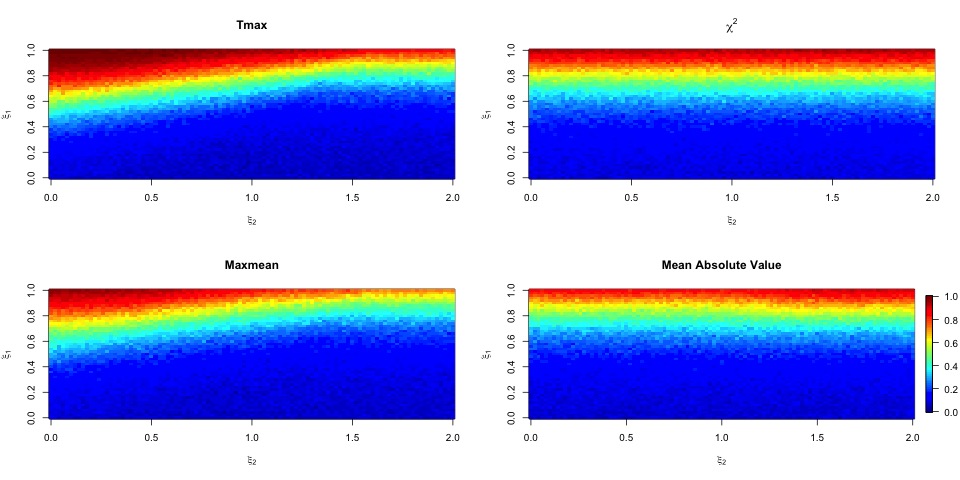}
\caption{Power comparison between different tests for signals simulated on pathway hsa00051.}
\label{fg:hsa00051}
\end{center}
\end{figure}

\begin{figure}[htbp]
\begin{center}
\includegraphics[width=\textwidth]{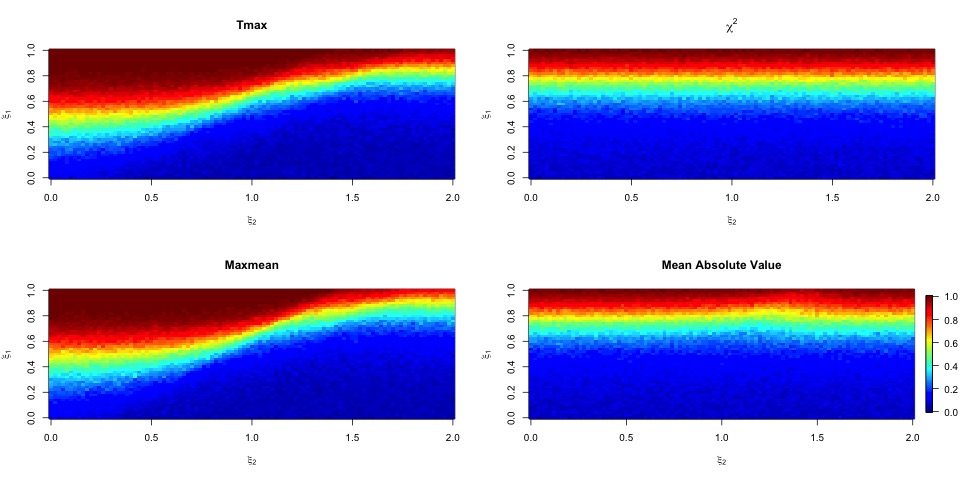}
\caption{Power comparison between different tests for signals simulated on pathway hsa00140.}
\label{fg:hsa00140}
\end{center}
\end{figure}

\begin{figure}[htbp]
\begin{center}
\includegraphics[width=\textwidth]{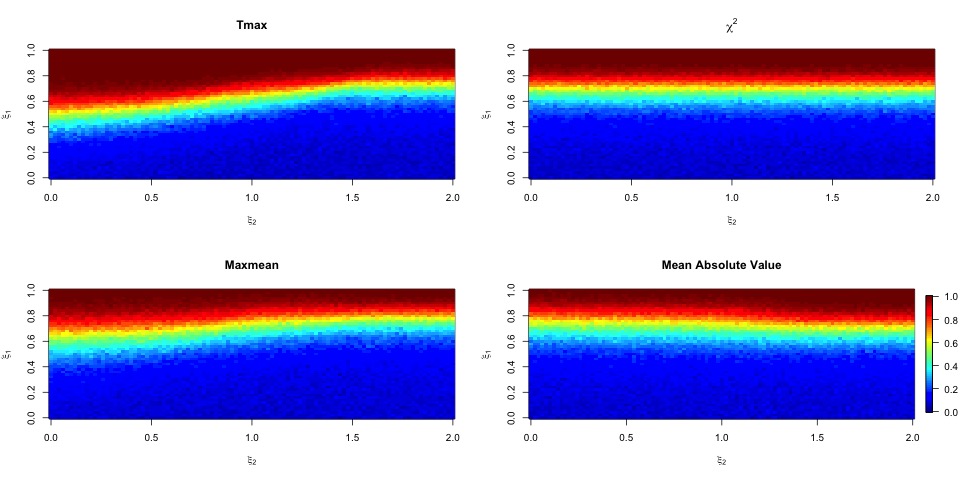}
\caption{Power comparison between different tests for signals simulated on pathway hsa00230.}
\label{fg:hsa00230}
\end{center}
\end{figure}

To further demonstrate the merits of our method, we now compare the performance of $T_{\max}$ based test with those based on several other commonly used statistics for gene set enrichment analysis: the maxmean statistic proposed by \cite{EfronTibshirani2007}; the mean of absolute values; and the $\chi^2$ statistic or the mean squares of the scores. To mimic realistic pathways, we simulated signals on three KEGG pathways: \verb+hsa00051+ with $33$ genes, \verb+hsa00140+ with $58$ genes, and \verb+hsa00230+ with $176$ genes. The three pathways were chosen to better illustrate the potential effect of the number of genes, and therefore compliment our asymptotic results. In addition, they are also among the pathways of interest in the NPC data example we shall present later. Detailed pathway information is accessible at \verb+http://www.genome.jp+. For each pathway, as before, we simulated signal $\bmu$ such that $\|\bmu\|^2=n^{\xi_1}$ and $\bmu^\top L\bmu=n^{\xi_2}$ where $n$ is the number of genes on the pathway and $L$ is the corresponding Laplacian. We calibrate the null distribution for each test statistic through 1000 Monte Carlo simulations. For each combination of $(\xi_1,\xi_2)$, we repeated the experiment in the same fashion for 500 times as before. The power of the test based on each test statistic is given in Figures \ref{fg:hsa00051}, \ref{fg:hsa00140} and \ref{fg:hsa00230} for each of the three pathways respectively. It is clear from these results that the $T_{\max}$ enjoys superior performance than the alternatives under all three settings.

\subsection{NPC data example}

Our final example is taken from a genome-wide expression study of nasopharyngeal carcinoma (NPC) (Sengupta et al. 2006). The goal of this study is to evaluate the association between host genes in NPC and a key gene in the infecting Epstein-Barr virus (EBV). The data, available from \texttt{allez} package in \verb+R+, has a total of $42346$ annotated probe sets. Following \cite{Newton2007} and \cite{NewtonWang2015}, a log-transformed Spearman correlation between viral gene EBNA1 and each human gene was used to evaluate the potential relationship between the viral gene and host genes. Six different gene set enrichment analysis methods were applied to this dataset: in addition to the four test statistics we considered before, Gene set enrichment analysis (GSEA) proposed by \cite{Subramanian2005} was also applied to these scores, as well as the  absolute value of these scores. We extracted pathway information for Homo sapiens (org code:hsa) in KEGG, leading to a collection of 301 pathways. We ignored genes on a particular pathway if they are not in our annotated probe sets. For each method, permutation test was applied to determine the $p$-value. To adjust for multiple comparison, we applied Benjamini-Hochberg procedure to control the false discovery rate at 0.1\%. The pathways that are identified by each method are given in Table~\ref{tb:signf}.

\begin{table}
\begin{center}
\begin{tabular}{c|p{14cm}}
\hline
Method & Pathways\\
\hline
\hline
$T_{\max}$ & hsa03013, hsa03030, hsa03040, hsa03430, hsa04010, hsa04014, hsa04020, hsa04024, hsa04060, hsa04062, hsa04064, hsa04110, hsa04514, hsa04612, hsa04620, hsa04630, hsa04640, hsa04650, hsa04660, hsa04662, hsa04664, hsa04713, hsa04740, hsa04744, hsa05166, hsa05169 \\
\hline
MeanAbs & hsa03013, hsa03030, hsa03040, hsa03430, hsa04110, hsa04114, hsa04612, hsa04640, hsa04650, hsa04660, hsa05169, hsa05340\\
\hline
Maxmean & hsa02010, hsa03008, hsa03013, hsa03030, hsa03040, hsa03430, hsa04020, hsa04060, hsa04062, hsa04064, hsa04080, hsa04110, hsa04261, hsa04380, hsa04514, hsa04612, hsa04630, hsa04640, hsa04650, hsa04660, hsa04662, hsa04672, hsa04713, hsa04740, hsa04925, hsa04940, hsa04970, hsa05320, hsa05321, hsa05330, hsa05332, hsa05340, hsa05414\\
\hline 
$\chi^2$ & hsa03013, hsa03030, hsa03430, hsa04110, hsa04114, hsa04612, hsa04640, hsa04650, hsa04660, hsa05166, hsa05169\\
\hline
GSEA & hsa00020, hsa00240, hsa00970, hsa00980, hsa03008, hsa03010, hsa03013, hsa03015, hsa03018, hsa03020, hsa03030, hsa03040, hsa03050, hsa03060, hsa03420, hsa03430, hsa04010, hsa04020, hsa04060, hsa04062, hsa04064, hsa04080, hsa04110, hsa04142, hsa04380, hsa04514, hsa04610, hsa04611, hsa04620, hsa04640, hsa04650, hsa04660, hsa04662, hsa04672, hsa04713, hsa04720, hsa04723, hsa04724, hsa04740, hsa04742, hsa04750, hsa04921, hsa04940, hsa04950, hsa04976, hsa05033, hsa05204, hsa05320, hsa05330, hsa05332, hsa05340, hsa05414\\
\hline 
GSEAAbs & hsa03013, hsa03030, hsa03040, hsa03430, hsa04110, hsa04612, hsa04640, hsa04650, hsa04940, hsa05320, hsa05330, hsa05332, hsa05340\\
\hline
\end{tabular}
\end{center}
\caption{Pathways identified by each method with false discovery rate controlled at 1\%.}
\label{tb:signf}
\end{table}%

To gain insights into the reliability of the lists of the pathways identified, we conducted another set of simulation to investigate the operating characteristics of these methods in a setting similar to the NPC data example. To this end, we simulated $42346$ scores to mimic the NPC data. Each score was simulated from a normal distribution with variance $\sigma^2=1.47^2$, which is the variance estimated from the NPC data. If a gene is not on any of the 26 pathways identified by the proposed method, its mean is set to zero. The means for genes on a pathway with Laplacian $L$, we fixed their mean as the same as a smoothed version of the observed scores from the NPC data:
$$
(I+\lambda L)^{-1} {\bf x},
$$
where ${\bf x}$ is the vector of observed scores for genes on the pathway from the NPC data, and $\lambda$ is taken to be the tuning parameter that maximizes $T_\lambda$. If a gene appears on multiple pathways, we average the means obtained from these pathways. We repeated the experiment for 1000 times and each time, we ran each of the six methods and recorded the lists of pathways they identified to have p-value smaller than 0.1\%. The power of each method, along with their false positive ratio, is summarized in Table \ref{tb:bootsimu}.

\begin{table}
\begin{center}
\begin{tabular}{c|c|c|c|c|c|c}
\hline
& graph &  maxmean & absmean & chisq &  gsea & gseaabs\\
\hline
\hline
hsa03013&0.989&1&0.889&0.877&0.994&0.374\\
hsa03030&0.889&0.998&0.837&0.889&0.954&0.724\\
hsa03040&1&1&1&1&0.973&0.81\\
hsa03430&0.539&0.983&0.595&0.539&0.91&0.657\\
hsa04010&0.974&1&0.314&0.216&0.825&0.037\\
hsa04014&0.807&0.998&0.068&0.079&0.713&0.005\\
hsa04020&0.854&1&0.657&0.486&0.968&0.227\\
hsa04024&0.856&0.998&0.219&0.142&0.868&0.061\\
hsa04060&0.999&1&0.962&0.96&1&0.723\\
hsa04062&0.998&1&0.336&0.307&0.989&0.116\\
hsa04064&0.745&0.946&0.289&0.43&0.632&0.048\\
hsa04110&0.995&1&0.51&0.489&0.941&0.123\\
hsa04514&0.978&1&0.898&0.875&0.938&0.258\\
hsa04612&0.999&0.989&0.944&0.951&0.276&0.358\\
hsa04620&0.839&0.829&0.084&0.087&0.867&0.058\\
hsa04630&1&1&0.349&0.552&0.97&0.035\\
hsa04640&0.98&1&0.975&0.98&0.94&0.578\\
hsa04650&0.998&1&0.485&0.733&0.998&0.287\\
hsa04660&0.861&0.989&0.263&0.246&0.59&0.016\\
hsa04662&0.845&0.825&0.185&0.182&0.448&0.046\\
hsa04664&0.444&0.883&0.057&0.069&0.402&0.023\\
hsa04713&0.994&0.978&0.247&0.086&0.588&0.008\\
hsa04740&0.926&0.997&0.353&0.313&0.995&0.364\\
hsa04744&0.536&0.7&0.099&0.084&0.415&0.194\\
hsa05166&0.848&0.924&0.744&0.653&0.257&0.176\\
hsa05169&0.998&0.33&0.985&0.985&0.042&0.57\\
\hline
False positive ratio &   0.075 & 0.186 & 0.027 & 0.030 & $0.079$ & $0.015$ \\
\hline
\end{tabular}
\end{center}
\caption{Comparison of power and false positive ratio between different methods based 1000 simulated datasets.}
\label{tb:bootsimu}
\end{table}%

%\section{Concluding Remarks}
%\label{sec:conclude}

\section{Proofs}
\label{sec:proof}

\begin{proof}[Proof of Theorem \ref{th:s2optimal}]
The main idea of the proof is to identify a set of carefully chosen $\bmu$s from $\Theta_G(\eta^2)$ such that $\|\bmu\|^2\ge c|V|^{1/2}$, and show that we can not distinguish them collectively from ${\bf 0}$. To this end, denote by $\bu\in \{\pm 1\}^{|V|}$ a vector of independent Rademacher random variables such that $\PP(u_i=+1)=\PP(u_i=-1)=1/2$. It is clear that
$$
\EE[\bu^\top L\bu]=\tr(L).
$$
Hereafter we write $L$ for $L(G)$ for short when no confusion occurs. By Hanson-Wright inequality \citep{HansonWright1971}, there exists a constant $C>0$ such that
$$
\PP\left\{|\bu^\top L\bu-\tr(L)|\ge t[\tr(L)]\right\}\le 2\exp\left(-C\min\left\{{t^2[\tr(L)]^2\over \|L\|_{\rm F}^2}, {t[\tr(L)]\over \|L\|}\right\}\right).
$$
In what follows, we shall use $C$ to denote a generic positive constant that may take different values at each appearance. Note that $L$ is a positive semidefinite matrix. Therefore,
$$
\|L\|_{\rm F}^2\le [\tr(L)]^2,\qquad {\rm and}\qquad \|L\|\le \tr(L).
$$
For any $t\ge 1$, we get
$$
\PP\left\{\bu^\top L\bu\le (t+1)\tr(L)\right\}\ge 1-\exp(-Ct).
$$
Because scaling does not change the rates of detection, we can assume without loss of generality that $\eta^2> \eta_{\max}^2(G)$. Then
$$
\PP\left\{|V|^{-1/2}\bu^\top L\bu\le \eta^2\right\}\ge 1-\exp(-C\eta^2/\eta_{\max}^2).
$$
Denote by $\calU$ the collection of all $\bu\in \{\pm 1\}^{|V|}$ such that
$$
\bu^\top L\bu\le |V|^{1/2}\eta^2.
$$
Then
$$
|\calU|\ge \left[1-\exp(-C\eta^2/\eta_{\max}^2)\right]2^{|V|}.
$$

Let $\PP_{\bmu}$ be the probability measure of $(X_v)_{v\in V}$ such that $X_v\sim N(\mu_v,1)$. Write
$$
\PP_1={1\over |\calU|}\sum_{\bu\in \calU} \PP_{\zeta |V|^{-1/4}\bu},
$$
for some $0\le \zeta\le 1$ to be specified later. Then for any test $\phi$, the sum of the probabilities of its two types of errors obeys
\begin{eqnarray*}
\sup_{\bu\in \calU} \left\{\EE_0 \phi + \EE_{|V|^{-1/4}\bu}(1-\phi)\right\}&\ge & \inf_\psi \sup_{\bu\in \calU} \left\{\EE_0 \psi + \EE_{|V|^{-1/4}\bu}(1-\psi)\right\}\\
&\ge& \inf_\psi {1\over |\calU|}\sum_{\bu\in \calU}\left\{\EE_0 \psi + \EE_{|V|^{-1/4}\bu}(1-\psi)\right\}\\
&=&1-{1\over 2}\|\PP_0-\PP_1\|_{\ell_1}.
\end{eqnarray*}
Recall that
$$
\|\PP_0-\PP_1\|_{\ell_1}^2\le \int{f_1^2\over f_0}-1,
$$
where $f_1$ and $f_0$ are the density functions corresponding to $\PP_0$ and $\PP_1$ respectively. 

It is not hard to see that
\begin{eqnarray*}
\int{f_1^2\over f_0}&=&{1\over |\calU|^2}\sum_{\bu_1,\bu_2\in \calU}\exp(\zeta^2|V|^{-1/2}\bu_1^\top\bu_2)\\
&\le&{1\over |\calU|^2}\sum_{\bu_1,\bu_2\in \{\pm 1\}^{|V|}}\exp(\zeta^2|V|^{-1/2}\bu_1^\top\bu_2)\\
&\le&\left[1-\exp(-C\eta^2/\eta_{\max}^2)\right]^{-2} \EE\exp(\zeta^2|V|^{-1/2}\bu_1^\top\bu_2),
\end{eqnarray*}
where the expectation is taken over two independent Radmacher random vectors $\bu_1$ and $\bu_2$. Note that
$$
\EE\exp(\zeta^2|V|^{-1/2}\bu_1^\top\bu_2)=\EE\exp[\zeta^2|V|^{-1/2}(2B-1)],
$$
where $B\sim {\rm Bin}(|V|,1/2)$. By Central Limit Theorem,
$$
\EE\exp[\zeta^2|V|^{-1/2}(2B-1)]\to \exp(\zeta^4/2).
$$
Therefore, when $|V|$ is large enough, for any test $\phi$,
\begin{eqnarray*}
\sup_{\bu\in \calU} \left\{\EE_0 \phi + \EE_{|V|^{-1/4}\bu}(1-\phi)\right\}&\ge& 1-{1\over 2}\left[1-\exp(-C\eta^2/\eta_{\max}^2)\right]^{-2}\sqrt{\exp(\zeta^4/2)-1}\\
&\ge&1-{\zeta^2\over 2}\left[1-\exp(-C\eta^2/\eta_{\max}^2)\right]^{-2},
\end{eqnarray*}
where in the second inequality we used the fact that $e^x\le 1+2x$ for any $x<1/2$. The desired claim then follows from the fact that $\|\zeta|V|^{-1/4}\bu\|^2=\zeta^2|V|^{1/2}$.
\end{proof}
\vskip 25pt

\begin{proof}[Proof of Theorem \ref{th:s1s2optimal}] Denote by $P_0$ the projection matrix from $\RR^{|V|}$ to the eigenspace of $L(G)$ corresponding to eigenvalue zero. It is not hard to see that, if $\bX\sim N(\bmu, I)$, then
$$
R\sim \chi^2_K(\|P_0\bmu\|^2).
$$
Observe that
$$
\bmu^\top L(G)\bmu\ge \eta_{\min}^2(G)\|P_0^\perp \bmu\|^2,
$$
so that
$$
\|P_0^\perp \bmu\|^2\le \eta^2/\eta_{\min}^2(G).
$$
Therefore,
$$
\|P_0\bmu\|^2=\|\bmu\|^2-\|P_0^\perp \bmu\|^2\ge \|\bmu\|^2-\eta^2/\eta_{\min}^2(G),
$$
which is of the same order as $\|\bmu\|^2$ if $\|\bmu\|^2\gg\eta^2/\eta_{\min}^2(G)$. The proof is then completed.
\end{proof}
\vskip 25pt

\begin{proof}[Proof of Theorem \ref{pr:ernull}]
We first note that an Erd\"os-R\'enyi graph $G_n$ is connected with probability tending to one suggesting that its Laplacian $L(G_n)$ has exactly one zero eigenvalue. Recall that $L(G_n)=D(G_n)-A(G_n)$ where $D(G_n)$ and $A(G_n)$ are $G_n$'s degree and adjacency matrices respectively. Applying random matrix theory, \cite{furedikomlos81} showed that the eigenvalues of $A(G_n)$ are $O_p(\sqrt{np})$ with the exception of the largest one. On the other hand, by Chernoff's bound, $\|D(G_n)-np\|=O_p(\sqrt{np})$. Thus, all nonzero eigenvalues of $L(G_n)$ are $np+O_p(\sqrt{np})$. In other words, we can write
$$
L(G_n)=J+\Delta,
$$
where
$$
J=(np)\left(I-{1\over n}\one\one^\top\right),
$$
and $\Delta$ is a symmetric matrix such that $\Delta\one =\zero$ and $\|\Delta\|=O_p(\sqrt{np})$.

Observe that
\begin{eqnarray*}
\bX^\top (I+\alpha L(G_n))^{-1}\bX&=&\bX^\top (I+\alpha J)^{-1}\left[I+\alpha(I+\alpha J)^{-1}\Delta)\right]^{-1}\bX\\
&=&\bX^\top (I+\alpha J)^{-1}\bX \cdot [1+O\left((np)^{-1}\|\Delta\|\right)].
\end{eqnarray*}
Similarly, we can show that
$$
\tr[(I+\alpha L(G_n))^{-1}]=\tr[(I+\alpha J)^{-1}]\cdot [1+O\left((np)^{-1}\|\Delta\|\right)],
$$
and
$$
\left\{\tr[(I+\alpha L(G_n))^{-2}]\right\}^{1/2}=\left\{\tr[(I+\alpha J)^{-2}]\right\}^{1/2}\cdot [1+O\left((np)^{-1}\|\Delta\|\right)].
$$
Therefore
$$
T_\alpha={\bX^\top (I+\alpha J)^{-1}\bX - \tr[(I+\alpha J)^{-1}]\over \left\{\tr[(I+\alpha J)^{-2}]\right\}^{1/2}}\cdot[1+O\left((np)^{-1}\|\Delta\|\right)],
$$
which implies that
\begin{equation}
\label{eq:equiv}
T_{\max}=\max_{\alpha\ge 0}T_\alpha =\max_{\alpha\ge 0}\left\{{\bX^\top (I+\alpha J)^{-1}\bX - \tr[(I+\alpha J)^{-1}]\over \left\{\tr[(I+\alpha J)^{-2}]\right\}^{1/2}}\right\}\cdot[1+O_p((np)^{-1/2})].
\end{equation}

On the other hand,
\begin{eqnarray*}
&&{\bX^\top (I+\alpha J)^{-1}\bX - \tr[(I+\alpha J)^{-1}]\over \left\{\tr[(I+\alpha J)^{-2}]\right\}^{1/2}}\\
&=&\left({n-1\over (1+np\alpha)^2}+1\right)^{-1/2}\left\{{1\over 1+np\alpha}\cdot \left[\|\bx_c\|^2-(n-1)\right]+(n\bar{x}^2-1)\right\}.
\end{eqnarray*}
Thus,
$$
\max_{\alpha\ge 0} \left\{{\bX^\top (I+\alpha J)^{-1}\bX - \tr[(I+\alpha J)^{-1}]\over \left\{\tr[(I+\alpha J)^{-2}]\right\}^{1/2}}\right\}=\max_{\theta\in [n^{-1/2},1]} \left\{\sqrt{1-\theta^2}\cdot W_1+\theta\cdot W_2\right\},
$$
where
$$
W_1={1\over \sqrt{n-1}}\left[\|\bx_c\|^2-(n-1)\right],\qquad {\rm and}\qquad W_2=n\bar{x}^2-1.
$$
Write
$$
h(\theta)=\sqrt{1-\theta^2}\cdot W_1+\theta\cdot W_2.
$$
It is clear that
$$
h'(\theta)=W_2-{\theta\over \sqrt{1-\theta^2}}\cdot W_1.
$$
By first order condition, we get
$$
\max_{\theta\in [n^{-1/2},1]} h(\theta)=\left\{\begin{array}{ll}(W_1^2+W_2^2)^{1/2}& {\rm if\ } W_1, W_2>0\\\max\left\{W_1,W_2\right\} & {\rm otherwise}\end{array}\right.
$$
It is not hard to see that
$$
W_1\to_d N(\sqrt{2}\delta_1^2,2),\qquad {\rm and}\qquad W_2\to_d \chi^2_1(\delta_2^2)-1.
$$
%It can then be derived that
%$$
%\max_{\alpha\ge 0} \left\{{\bX^\top (I+\alpha J)^{-1}\bX - \tr[(I+\alpha J)^{-1}]\over \left\{\tr[(I+\alpha J)^{-2}]\right\}^{1/2}}\right\}\to_d\left\{\begin{array}{ll}(2Y_1^2+(Y_2^2-1)^2)^{1/2}& {\rm if\ } Y_1>0, Y_2^2>1\\\max\left\{\sqrt{2}Y_1,Y_2^2-1\right\} & {\rm otherwise}\end{array}\right..
%$$
This, together with (\ref{eq:equiv}), implies (\ref{eq:knHa}).
\end{proof}
\vskip 25pt

\begin{proof}[Proof of Theorem \ref{th:er}] By Theorem \ref{pr:ernull}, $\varphi$ is consistent for testing against any effect $\bmu$ such that
\begin{equation}
\label{eq:complete}
{1\over n^{1/2}}\|\bmu_c\|^2+n\bar{\mu}^2\to \infty,\qquad {\rm as}\quad n\to\infty.
\end{equation}
In addition, as shown in the proof of Theorem \ref{pr:ernull},
$$
L(G_n)=(np)J+\Delta
$$
such that $\Delta\one=\zero$ and $\|\Delta\|=O_p(\sqrt{np})$. Thus,
$$
\bmu^\top L(G_n)\bmu=(np)\bmu^\top J\bmu+O_p(\|\bmu_c\|^2\sqrt{np})=\|\bmu_c\|^2(np+O_p(\sqrt{np})),
$$
which implies that $\|\bmu_c\|^2\gg n^{1/2}$ if $\eta^2\gg n^{3/2}$. Together with the fact that
$$
\|\bmu\|^2=\|\bmu_c\|^2+n\bar{\mu}^2,
$$
this immediately implies that when $\eta^2\gg n^{3/2}$, $\beta(\varphi;\bmu)\to 0$ if $\|\bmu\|^2\gg n^{1/2}$; and on the other hand, when $\eta^2=O(n^{3/2})$, $\beta(\varphi;\bmu)\to 0$ if $\|\bmu\|^2=\eta^2/n+\omega(1)$. This completes the proof of the first statement.

We now show that this indeed is the best one can do. Note that
$$\tr(L(G_n))=n(n-1)p(1+o_p(1)).$$
The lower bound for the case when $\eta^2=\Omega(n^{3/2})$ then follows immediately from Theorem \ref{th:s2optimal}. Similarly, the lower bound for the case when $\eta=O(n^{1/2})$ follows from Theorem \ref{th:s1s2optimal} since the minimum nonzero eigenvalue of $L(G_n)$ is of the form $np+O_p(\sqrt{np})$. It remains to treat the case when $n^{1/2}\ll \eta\ll n^{3/4}$.

To this end, let $\bw_1,\ldots,\bw_{n-1}$ be an (arbitrary) orthogonal basis of the linear subspace
$$\{\bx\in \RR^{|V|}: \bx^\top \one=\zero\}.$$
Write $k_\ast=\lfloor \eta^4/n^2\rfloor$. For any $\bu\in \{\pm 1\}^{k_\ast}$, denote by $\PP_{\bu}$ the probability measure of $\bX=(X_v)_{v\in V}$ such that $\bX\sim N(\bmu, I_n)$, where
\begin{equation}
\label{eq:defmu}
\bmu=\zeta\eta(nk_\ast)^{-1/2}\sum_{k=1}^{k_\ast} u_k \bw_k,
\end{equation}
for some $\zeta<1$ to be specified later. It is not hard to see that, with this choice,
$$
\bmu^\top L(G_n)\bmu=\zeta^2\eta^2(nk_\ast)^{-1}\sum_{k=1}^{k_\ast} \lambda_k(G_n)=\zeta^2\eta^2p(1+o_p(1)),
$$
indicating $\bmu\in \Theta_{G_n}(\eta^2)$ with probability tending to one. As before, denote by $\calU$ the collection of $\bu$ such that the corresponding $\bmu$ as defined by (\ref{eq:defmu}) belongs to $\Theta_{G_n}(\eta^2)$. Then $|\calU|/2^{k_\ast}\to_p 1$. Write
$$
\PP_1={1\over |\calU|}\sum_{\bu\in \calU} \PP_{\bu}.
$$
Following the same calculation as before, it suffices to show that $\int(f_1^2/ f_0)$ can be made arbitrarily close to $1$. Recall that
$$
\int{f_1^2\over f_0}= \EE\exp[\zeta^2\eta^2(nk_\ast)^{-1}\bu_1^\top \bu_2],
$$
where the expectation is taken over two independent random vectors $\bu_1$ and $\bu_2$ uniformly sampled from $\calU$. Following a similar argument as that of Theorem \ref{th:s2optimal}, we can derive that
$$
\EE\exp[\zeta^2\eta^2(nk_\ast)^{-1}\bu_1^\top \bu_2]\to \exp(\zeta^4/2).
$$
By taking $\zeta$ small enough, we can ensure that any test is powerless in testing against $\bmu$ of the form (\ref{eq:defmu}) with $\bu\in\calU$. The proof is then completed by noting that
$$
\|\bmu\|^2\le {\zeta^2\eta^2\over n},
$$
for any $\bu\in \calU$.
\end{proof}
\vskip 25pt

\begin{proof}[Proof of Theorem \ref{th:power}]
For brevity, we omit the dependence of the Laplacian matrix $L$ on $G$ and write $n:=|V|$ throughout the proof. We first prove the first statement. To this end, let
$$
\alpha_\ast=\argmax_{\alpha}\left\{\|H(\alpha)\|_{\rm F}^{-1}\bmu^\top H(\alpha) \bmu\right\},
$$
where
$$
H(\alpha)=(I+\alpha L)^{-1}.
$$
Of course, the maximizer may not be unique, in which case, $\alpha_\ast$ is chosen arbitrarily among the maximizers. By Hanson-Wright inequality \citep{HansonWright1971},
$$
\PP\left\{|T_{\alpha_\ast}-\|H(\alpha_\ast)\|_{\rm F}^{-1}\bmu^\top H(\alpha_\ast) \bmu|\ge t\right\}\le 2\exp\left(-C\min\left\{t^2, {t\|H(\alpha_\ast)\|_{\rm F}}\right\}\right),
$$
This immediately implies that
$$
\PP\left\{T_{\alpha_\ast}\le {1\over 2}\|H(\alpha_\ast)\|_{\rm F}^{-1}\bmu^\top H(\alpha_\ast) \bmu\right\}\le 2\exp\left(-C\min\left\{\|H(\alpha_\ast)\|_{\rm F}^{-2}\left(\bmu^\top H(\alpha_\ast) \bmu\right)^2, \bmu^\top H(\alpha_\ast) \bmu\right\}\right).
$$
It is therefore clear that
$$
T_{\max}\ge T_{\alpha_\ast}\gg \log\log n,
$$
with probability tending to one, by assumption (\ref{eq:altercond}). It now suffices to show that under $H_0$,
$$
T_{\max}=O_p(\log\log n).
$$

With slight abuse of notation, let $0=\rho_0<\rho_1<\cdots<\rho_N$ be the distinct eigenvalues of $L$ and $n_k$ be the multiplicity of $\rho_k$.
Write $\bw_\alpha=(w_{\alpha,1},\ldots,w_{\alpha, N})^\top$ where
$$
w_{\alpha, k}={\sqrt{n_k}\over 1+\alpha\rho_k},\qquad k=0,1,\ldots, N.
$$
Then, under $H_0$, $T_{\max}$ follows the same distribution as
$$
\sup_{\alpha\ge 0} \left\{\|\bw_\alpha\|^{-1}\sum_{k=0}^N w_{\alpha, k} Y_k\right\},
$$
where $Y_k\sim n_k^{-1/2}(\chi^2_{n_k}-n_k)$, $k=0,1,\ldots, N$, are independent random variables. Note that for any $0<\alpha_{\min}<\alpha_{\max}<\infty$,
\begin{eqnarray*}
&&\sup_{\alpha\ge 0} \left\{\|\bw_\alpha\|^{-1}\sum_{k=0}^N w_{\alpha, k} Y_k\right\}\\
&=&\max\biggl\{\sup_{\alpha\in [0,\alpha_{\min}]} \{\|\bw_\alpha\|^{-1}\sum_{k=0}^N w_{\alpha, k} Y_k\},\sup_{\alpha\in (\alpha_{\min},\alpha_{\max})} \{\|\bw_\alpha\|^{-1}\sum_{k=0}^N w_{\alpha, k} Y_k\},\\
&&\hskip 50pt \sup_{\alpha>\alpha_{\max}} \{\|\bw_\alpha\|^{-1}\sum_{k=0}^N w_{\alpha, k} Y_k\}\biggr\}
\end{eqnarray*}
We now treat the three terms on the righthand side separately with appropriately chosen $\alpha_{\min}$ and $\alpha_{\max}$.

\paragraph{Small $\alpha$s.} We first consider the case when $\alpha$ is small in that
$$\alpha\le\alpha_{\min}:={1\over [\tr(L^2)]^{1/2}\log (N+1)}.$$
Observe that, if $\alpha\le \alpha_{\min}$, then $\alpha\rho_k\le 1$ for $k=1,\ldots, N$. Thus,
$$
\|\bw_\alpha\|^2=\sum_{k=0}^N{n_k\over (1+\alpha \rho_k)^2}\ge {n\over 4}.
$$
We then get
\begin{eqnarray*}
\sup_{\alpha\in [0,\alpha_{\min}]}\left\{\|\bw_\alpha\|^{-1}\sum_{k=0}^N w_{\alpha, k} Y_k\right\}&\le& \|\bw_{\alpha_{\min}}\|^{-1}\left(\left|\sum_{k=0}^N \sqrt{n_k}Y_k\right|+\sup_{\alpha\in [0,\alpha_{\min}]}\left|\sum_{k=0}^N (w_{\alpha, k}-\sqrt{n_k}) Y_k\right|\right)\\
&\le& \left(\sqrt{4\over n}\left|\sum_{k=0}^N \sqrt{n_k}Y_k\right|+\|\bw_{\alpha_{\min}}\|^{-1}\sup_{\alpha\in [0,\alpha_{\min}]}\left|\sum_{k=0}^N (w_{\alpha, k}-\sqrt{n_k}) Y_k\right|\right)\\
&\le& O_p(1)+\|\bw_{\alpha_{\min}}\|^{-1}\sup_{\alpha\in [0,\alpha_{\min}]}\left|\sum_{k=0}^N (w_{\alpha, k}-\sqrt{n_k}) Y_k\right|.
\end{eqnarray*}
where the last inequality follows from Markov inequality and the fact that
$$
\EE\left|\sum_{k=0}^N \sqrt{n_k}Y_k\right|^2=\sum_{k=0}^N n_k\EE Y_k^2=2n.
$$

Now note that for any $\alpha\in [0,\alpha_{\min}]$, $\alpha\rho_k\le 1$ so that
$$
\left|\sum_{k=0}^N (w_{\alpha, k}-\sqrt{n_k}) Y_k\right|\le \left(\sum_{k=1}^N {\sqrt{n_k}\alpha\rho_k\over 1+\alpha\rho_k}\right)\max_{1\le k\le N}|Y_k|.
$$
By Cauchy-Schwartz inequality,
\begin{eqnarray*}
\left|\sum_{k=0}^N (w_{\alpha, k}-\sqrt{n_k}) Y_k\right|&\le&\alpha\|\bw_\alpha\|\left(\sum_{k=1}^N n_k\rho_k^2\right)^{1/2}\max_{1\le k\le N}|Y_k|\\
&=&\alpha\|\bw_\alpha\|[\tr(L^2)]^{1/2}\max_{1\le k\le N}|Y_k|.%=O_p(\sqrt{n}),
\end{eqnarray*}
By the choice of $\alpha_{\min}$, together with the fact that
$$
\max_{1\le k\le N}|Y_k|=O_p(\log N),
$$
we get
$$
\sup_{\alpha\in [0,\alpha_{\min}]}\left\{\|\bw_\alpha\|^{-1}\sum_{k=0}^N w_{\alpha, k} Y_k\right\}=O_p(1).
$$

\paragraph{Large $\alpha$s} Next we consider the case when $\alpha$ is large in that
$$
\alpha\ge \alpha_{\max}:={\log (N+1)\over \sqrt{n_0}}\sum_{k=1}^N{\sqrt{n_k}\over \rho_k},
$$
where $n_0=K$ is the number of connected components of $G$.

It is clear that $\|\bw_{\alpha}\|\ge w_{\alpha, 0}=\sqrt{n_0}$. Thus,
\begin{eqnarray*}
\|\bw_\alpha\|^{-1}\sum_{k=0}^N w_{\alpha, k} Y_k&\le& w_{\alpha,0}^{-1}\left|\sum_{k=0}^N w_{\alpha, k} Y_k\right|\\
&\le&|Y_0|+w_{\alpha,0}^{-1}\left|\sum_{k=1}^N w_{\alpha, k} Y_k\right|\\
&\le&O_p(1)+w_{\alpha,0}^{-1}\left(\sum_{k=1}^N w_{\alpha, k}\right)\max_{1\le k\le N}\left|Y_k\right|.
\end{eqnarray*}
Recall that, for any $\alpha\ge \alpha_{\max}$,
$$
\sum_{k=1}^N w_{\alpha, k}=\sum_{k=1}^N {\sqrt{n_k}\over 1+\alpha\rho_k}\le \sum_{k=1}^N {\sqrt{n_k}\over \alpha\rho_k}\le {\sqrt{n_0}\over \log (N+1)}.
$$
Together with the fact that
$$
\max_{1\le k\le N}\left|Y_k\right|=O_p(\log N),
$$
we get
$$
\sup_{\alpha\ge \alpha_{\max}}\left\{\|\bw_\alpha\|^{-1}\sum_{k=1}^n w_{\alpha, k} Y_k\right\}=O_p(1).
$$

\paragraph{Intermediate $\alpha$s.} Finally, we treat the case when $\alpha\in (\alpha_{\min},\alpha_{\max})$. To this end, we write $\alpha_m=2^{m-1}\alpha_{\min}$, for $m=1,\ldots, \lceil \log_2(\alpha_{\max}/\alpha_{\min})\rceil$. It is clear that
\begin{eqnarray*}
&&\sup_{\alpha\in(\alpha_{\min}, \alpha_{\max})}\left\{\|\bw_\alpha\|^{-1}\sum_{k=0}^N w_{\alpha, k} Y_k\right\}\\
&\le&\max_{1\le m\le M}\sup_{\alpha\in[\alpha_{m}, \alpha_{m+1})}\left\{\|\bw_\alpha\|^{-1}\sum_{k=0}^N w_{\alpha, k} Y_k\right\}\\
&\le&\max_{1\le m\le M}\left\{\sup_{\alpha\in[\alpha_{m}, \alpha_{m+1})}\|\bw_\alpha\|^{-1}\sum_{k=0}^N w_{\alpha, k} Y_k-\|\bw_{\alpha_m}\|^{-1}\sum_{k=0}^N w_{\alpha_m, k} Y_k\right\}\\
&&\hskip 50pt +\max_{1\le m\le M}\left\{\|\bw_{\alpha_m}\|^{-1}\sum_{k=0}^N w_{\alpha_m, k} Y_k\right\},
\end{eqnarray*}
where $M=\lfloor \log_2(\alpha_{\max}/\alpha_{\min})\rfloor$.

%Observe that $\|\bw_\alpha\|$ is a decreasing function of $\alpha$, so that
%\begin{eqnarray*}
%&&\sup_{\alpha\in[\alpha_{m}, \alpha_{m+1})}\|\bw_\alpha\|^{-1}\sum_{k=0}^N w_{\alpha, k} Y_k-\|\bw_{\alpha_m}\|^{-1}\sum_{k=0}^N w_{\alpha_m, k} Y_k\\
%&\le&\|\bw_{\alpha_m}\|^{-1}\left[\sup_{\alpha\in[\alpha_{m}, \alpha_{m+1})}\sum_{k=1}^N (w_{\alpha, k}-w_{\alpha_m,k}) Y_k\right].
%\end{eqnarray*}
Note that, for any $\alpha_m\le \alpha<\beta<\alpha_{m+1}$,
$$
0\le w_{\alpha,k}-w_{\beta,k}={\sqrt{n_k}(\beta-\alpha)\rho_k\over (1+\alpha\rho_k)(1+\beta\rho_k)}\le(1-\alpha/\beta) w_{\alpha,k},
$$
which implies that
\begin{equation}
\label{eq:norm}
d_2(\alpha,\beta):=\left\|{\bw_{\alpha}\over \|\bw_\alpha\|}-{\bw_{\beta}\over \|\bw_\beta\|}\right\|\le 2(1-\alpha/\beta).
\end{equation}
Moreover, for any $\alpha_m\le \alpha<\beta<\alpha_{m+1}$, 
$$
0\le {1\over 1+\alpha\rho_k}-{1\over 1+\beta\rho_k}\le w_{\alpha,k}-w_{\beta,k}\le(1-\alpha/\beta) w_{\alpha,k}.
$$
This suggests that
$$
d_\infty(\alpha,\beta):=\max_{k}\left|{1\over \|\bw_\alpha\|(1+\alpha\rho_k)}-{1\over \|\bw_\beta\|(1+\beta\rho_k)}\right|\le 2(1-\alpha/\beta).
$$
On the other hand, by Hanson-Wright inequality \citep{HansonWright1971}, there exists a constant $C>0$ such that for any $\alpha>0$,
\begin{equation}
\label{eq:hw}
\PP\left\{\left|\|\bw_\alpha\|^{-1}\sum_{k=0}^N w_{\alpha, k} Y_k-\|\bw_\beta\|^{-1}\sum_{k=0}^N w_{\beta, k} Y_k\right|\ge t\right\}\le 2\exp\left(-C\min\left\{{t^2\over d^2_2(\alpha,\beta)},{t\over d_\infty(\alpha,\beta)}\right\}\right).
\end{equation}
We can apply a generic chaining argument to bound the supreme over $\alpha\in [\alpha_m,\alpha_{m+1})$:
$$
\PP\left\{\sup_{\alpha\in[\alpha_{m}, \alpha_{m+1})}\left|\|\bw_\alpha\|^{-1}\sum_{k=0}^N w_{\alpha, k} Y_k-\|\bw_{\alpha_m}\|^{-1}\sum_{k=0}^N w_{\alpha_m, k} Y_k\right|\ge t\right\}\le C_1\exp(-C_2t)
$$
for some constant $C1,C_2>0$. See, e.g., Theorem 2.2.23 of \cite{Talagrand2014}.

Now an application of union bounds over $m$ yields,
$$
\sup_{\alpha\in(\alpha_{\min}, \alpha_{\max})}\left\{\|\bw_\alpha\|^{-1}\sum_{k=0}^N w_{\alpha, k} Y_k\right\}=O_p\left(\log\log(\alpha_{\max}/\alpha_{\min})\right)=O_p(\log\log n),
$$
where the last equality follows from the assumption on $\rho_1=\lambda_{\min}$ and the fact that $\|L\|_{\rm F}^2\le 2\|D(G)\|_{\rm F}^2\le 2n^3$. This then implies the consistency of $\varphi_T$ over all $\bmu$ that satisfies (\ref{eq:altercond}). 

Now, to prove (\ref{eq:bdgeneral}), it suffices to show that it implies (\ref{eq:altercond}). Let $\bX\sim N(\bmu,I)$ for some $\bmu$ obeying (\ref{eq:bdgeneral}). Note that
$$
\bX^\top (I+\alpha L)^{-1}\bX=\bmu^\top (I+\alpha L)^{-1}\bmu + 2\bmu^\top(I+\alpha L)^{-1}\beps +\beps^\top (I+\alpha L)^{-1}\beps,
$$
where $\beps=\bX-\bmu$. We can write
$$
T_\alpha=T_\alpha^{(1)}+T_\alpha^{(2)}+T_\alpha^{(3)},
$$
where
\begin{eqnarray*}
T_\alpha^{(1)}&=&{\bmu^\top (I+\alpha L)^{-1}\bmu\over \{\tr[(I+\alpha L)^{-2}]\}^{1/2}},\\
T_\alpha^{(2)}&=&{2\bmu^\top(I+\alpha L)^{-1}\beps\over \{\tr[(I+\alpha L)^{-2}]\}^{1/2}},\\
T_\alpha^{(3)}&=&{\beps^\top (I+\alpha L)^{-1}\beps-\tr[(I+\alpha L)^{-1}]\over \{\tr[(I+\alpha L)^{-2}]\}^{1/2}}.
\end{eqnarray*}
Observe that
$$
(I+\alpha L)^{-1}\succeq 1-\alpha L.
$$
Therefore,
$$
T_\alpha^{(1)}\ge{\bmu^\top \bmu -\alpha \bmu^\top L\bmu\over \{\tr[(I+\alpha L)^{-2}]\}^{1/2}}\ge {\bmu^\top\bmu -\alpha \eta^2\over \{\tr[(I+\alpha L)^{-2}]\}^{1/2}}.
$$
Taking $\alpha= x_\ast(G, \eta^2)/(2\eta^2)$ yields
$$
T_\alpha^{(1)}\ge{\bmu^\top \bmu\over 2\{\tr[(I+\alpha L)^{-2}]\}^{1/2}}.
$$
Recall that (\ref{eq:bdgeneral}) means
$$\|\bmu\|^2/x_\ast(G, \eta^2)\to \infty,$$
and (\ref{eq:master}) implies that
$$
x_\ast^2(G, \eta^2)=(\log\log n)\cdot\tr[(I+2\alpha L)^{-2}]\le {\log\log n\over 4}\tr[(I+\alpha L)^{-2}].
$$
We have $T_\alpha^{(1)}\gg \log\log n$ as a result.

On the other hand,
$$
T_\alpha^{(2)}\sim N\left(0,{4\bmu^\top (I+\alpha L)^{-2} \bmu\over \tr[(I+\alpha L)^{-2}]}\right).
$$
Note that
$$
{4\bmu^\top (I+\alpha L)^{-2} \bmu\over \tr[(I+\alpha L)^{-2}]}\le {4\bmu^\top \bmu\over \tr[(I+\alpha L)^{-2}]}.
$$
Therefore,
$$
T_\alpha^{(2)}=O_p\left({\|\bmu\|\over \left\{\tr[(I+\alpha L)^{-2}]\right\}^{1/2}}\right)=O_p\left(\sqrt{T_\alpha^{(1)}}\right)=o_p(T_\alpha^{(1)}).
$$
Together with the fact that $T_\alpha^{(3)}=O_p(1)$, we get
$$T_{\max}\ge T_\alpha\gg \log\log n,$$
with probability tending to one. This, together with the fact that $T_{\max}=O_p(\log\log n)$ under $H_0$, implies the consistency of $\varphi_T$.
\end{proof}
\vskip 25pt

\begin{proof}[Proof of Theorem \ref{th:star}]
We now show that $\varphi_T$ is consistent in testing against any $\bmu\in \Theta_{S_n}(\eta^2)$ such that $\|\bmu\|^2\gg r^2_{S_n}(\eta^2)$. It is clear that the leading eigenvector of $L(S_n)$ is 
$$
\bw_1:={1\over \sqrt{n(n-1)}}(n-1,-1,\ldots,-1)^\top,
$$
and the eigenvector corresponds to $\lambda_n(L(S_n))=0$ is $\bw_n=\one/\sqrt{n}$. Denote by $Y_1=\bX^\top \bw_1$ and $Y_2=\bX^\top\bw_n$. Let $\bZ=P \bX$ where $P$ is the projection matrix from $\RR^{|V|}$ to the eigenspace corresponding to eigenvalue one, i.e., the linear subspace of $\RR^n$ perpendicular to the linear space spanned by $\bw_1$ and $\bw_n$. It is not hard to see that
$$
\bX^\top (I+\lambda L(S_n))^{-1}\bX=(1+n\lambda)^{-1}Y_1^2+Y_2^2+(1+\lambda)^{-1}\|\bZ\|^2.
$$
Thus,
$$
T_\alpha={(1+n\alpha)^{-1}(Y_1^2-1)+(Y_2^2-1)+(1+\alpha)^{-1}(\|\bZ\|^2-(n-2))\over \left[(1+n\alpha)^{-2}+1+(n-2)(1+\alpha)^{-2}\right]^{1/2}}.
$$
Write
$$
Z={1\over \sqrt{n-2}}(\|\bZ\|^2-(n-2)).
$$
It is clear that, under $H_0$, $Y_1,Y_2\sim N(0,1)$ and $Z\to_d N(0,2)$, so that
$$
T_{\max}\le |Y_1^2-1|+|Y_1^2-1|+|Z|=O_p(1).
$$
It now suffices to show that if $\bX\sim N(\bmu,I)$ for any $\bmu\in \Theta_{S_n}(\eta^2)$ such that $\|\bmu\|^2\gg r^2_{S_n}(\eta^2)$, then $T_{\max}\to_p \infty$.

We first consider the case when $\eta^2\le 1$. Recall that
$$
\|\bmu\|^2=n\bar{\mu}^2+\|\bmu_c\|^2\le n\bar{\mu}^2+\bmu^\top L(S_n)\bmu,
$$
so that
$$
n\bar{\mu}^2\ge \|\bmu\|^2-\eta^2\ge \|\bmu\|^2-1.
$$
Observe that $Y_2\sim N(\sqrt{n}\bar{\mu},1)$. We have
$$
T_{\max}\ge T_\infty=Y_2^2-1\to_p +\infty
$$
as long as $\|\bmu\|^2\gg 1$.

Similarly, if $1\le \eta^2\le n^{1/2}$, then
$$
n\bar{\mu}^2\ge \|\bmu\|^2-\eta^2\to \infty
$$
as long as $\|\bmu\|^2\gg \eta^2$, so that $\varphi_T$ is consistent if $\|\bmu\|^2\gg \eta^2$.

Finally, the case when $\eta^2>n^{1/2}$ follows immediately from the facts that $T_{\max}\ge T_0$ and $T_0\sim \chi^2_n(\|\mu\|^2)$.

%Finally, we look at the case when $\eta^2>n^{1/2}$. Recall that
%$$
%\|\bmu\|^2=n\bar{\mu}^2+(\bmu^\top \bw_1)^2+\|P\bmu\|^2.
%$$
%Therefore,
%$$
%\max\{n\bar{\mu}^2,(\bmu^\top \bw_1)^2,\|P\bmu\|^2\}\ge {1\over 3}\|\bmu\|^2\gg n^{1/2}.
%$$
%As before, $n\bar{\mu}^2\gg n^{1/2}$ immediately implies that $T_{\max}\ge T_\infty\to_p\infty$. If $\|P\bmu\|^2\gg n^{1/2}$, then
%$$
%T_{\max}\ge T_0=n^{-1/2}\left[(Y_1^2-1)+(Y_2^2-1)+(n-2)^{1/2}Z\right]\ge n^{-1/2}\left[(n-2)^{1/2}Z-2\right].
%$$
%Note that
%$$
%\EE Z=(n-2)^{-1/2}\|P\bmu\|^2\to \infty.
%$$
%We know that $T_{\max}\to_p\infty$. If $(\bmu^\top \bw_1)^2\gg n^{1/2}$ and $\|P\bmu\|^2=O(n^{1/2})$, then
%\begin{eqnarray*}
%T_{\max}&\ge& T_0\\
%&=&n^{-1/2}\left[(Y_1^2-1)+(Y_2^2-1)+(n-2)^{1/2}Z\right]\\
%&\ge& n^{-1/2}\left((\bmu^\top \bw_1)^2-2\right)+n^{-1/2}\left(Y_2^2-(\bmu^\top \bw_1)^2\right)+\left(1-{2\over n}\right)^{1/2} Z.
%\end{eqnarray*}
%Observe that $Y_2\sim N(\bmu^\top \bw_1,1)$. We have
%$$
%\EE \left[Y_2^2-(\bmu^\top \bw_1)^2\right]^2=6(\bmu^\top \bw_1)^2+3.
%$$
%Therefore,
%$$
%n^{-1/2}\left(Y_2^2-(\bmu^\top \bw_1)^2\right)=O_p\left(n^{-1/2}|\bmu^\top \bw_1|\right)=o_p\left(n^{-1/2}(\bmu^\top \bw_1)^2\right),
%$$
%where the last equality follows from the fact that $(\bmu^\top \bw_1)^2\to \infty$. On the other hand,
%$$
%\EE Z^2={2\over n-2}\left(n-2+2\|P\bmu\|^2\right)=O(1),
%$$
%where we used the fact that $\|P\bmu\|^2=O(n^{1/2})$. Thus,
%$$
%\left(1-{2\over n}\right)^{1/2} Z=O_p(1)=o_p\left(n^{-1/2}(\bmu^\top \bw_1)^2\right).
%$$
%In summary, we have
%$$
%T_{\max}\ge n^{-1/2}(\bmu^\top \bw_1)^2\cdot [1+o_p(1)]\to \infty.
%$$

We now show that $r^2_{S_n}(\eta^2)$ indeed is the optimal detection boundary. The optimality when $\eta^2=O(1)$ or $\eta^2\ge n^{1/2}$ follows from Theorems \ref{th:s1s2optimal} and \ref{th:s2optimal} respectively. The case when $1\ll \eta^2\ll n^{1/2}$ can be treated in an identical fashion as Theorem \ref{th:er}. The only exception is now we take $\bw_2,\ldots,\bw_{n-1}$ to be an orthonormal basis of the eigenspace corresponding to eigenvalue one, and in defining $\bmu$ as in Equation (\ref{eq:defmu}), we sum from $k=2$.
\end{proof}
\vskip 25pt

\begin{proof}[Proof of Theorem \ref{th:cycle}]
Denote by $\{\bw_1,\ldots,\bw_n\}$ the eigenvectors corresponding to the eigenvalues $\lambda_1>\lambda_2>\cdots>\lambda_n=0$ of $L(C_n)$ sorted in decreasing order. Write $\eta_j=2^{5j/4}n^{-1}(\log\log n)^{-1/4}$ and $\bc_j=(c_{j1},\ldots,c_{jn})^\top$ where
\begin{equation}
\label{eq:defcpath}
c_{jk}=\zeta\cdot\left\{\begin{array}{ll}(\log\log n)^{1/4}(n\eta_j)^{-1/5}& {\rm if\ } (n\eta_j)^{4/5}/2\le n-k\le (n\eta_j)^{4/5}\\ 0& {\rm otherwise}\end{array}\right.,
\end{equation}
for some $\zeta>0$ to be determined later. Denote by
$$
\bmu_{j,\bu}=\sum_{k=1}^n c_{jk}u_k \bw_k,
$$
where $\bu\in \{\pm 1\}^n$. 
Recall that
$$
\bmu_{j,\bu}^\top L(C_n)\bmu_{j,\bu}=\sum_{(n\eta_j)^{4/5}/2\le n-k\le (n\eta_j)^{4/5}} \lambda_k c_{jk}^2.
$$
It is well known that there exists a constant $c>0$ such that
$$
\lambda_k\le c\cdot{(n-k)^2\pi^2\over n^2}.
$$
Thus,
$$
\bmu_{j,\bu}^\top L(C_n)\bmu_{j,\bu}\le {c(n\eta_j)^{8/5}\pi^2\over n^2}\cdot {1\over 2}(n\eta_j)^{4/5}\cdot\zeta^2(n\eta_j)^{-2/5}(\log\log n)^{1/2}={c\over 2}\zeta^2\pi^2\eta_j^2(\log\log n)^{1/2}.
$$
Taking $\zeta\le \sqrt{2/c\pi^2}$ ensures that
$$
\bmu_{j,\bu}^\top L(C_n)\bmu_{j,\bu}\le\eta_j^2(\log\log n)^{1/2}=4^{5j/4}n^{-2}=:\tilde{\eta}_j^2.
$$
On the other hand,
\begin{eqnarray*}
\|\bmu_{j,\bu}\|^2&=&\sum_{(n\eta_j)^{4/5}/2\le n-k\le (n\eta_j)^{4/5}}  c_{jk}^2\\
&\le& {1\over 2}(n\eta_j)^{4/5}\cdot\zeta^2(n\eta_j)^{-2/5}(\log\log n)^{1/2}\\
&=&{\zeta^2\over 2}(n\eta_j)^{2/5}(\log\log n)^{1/2}\\
&=&{\zeta^2\over 2}2^{j/2}(\log\log n)^{2/5}\\
&=&{\zeta^2\over 2}(n\tilde{\eta}_j\log\log n)^{2/5}.
\end{eqnarray*}

Now write
$$
\PP_1={1\over J}\sum_{j=1}^{J}\PP_{\bc_j},
$$
where $J=\lceil (5/4)\log_2 n\rceil$ and 
$$
\PP_{\bc_j}={1\over 2^n}\sum_{\bu\in \{\pm 1\}^n}\bmu_{j,\bu}.
$$
It is not hard to see that
$$
\int{f_1^2\over f_0}={1\over J}\sum_{j=1}^J\EE\exp(\bu_1^\top {\rm diag}(c_{j1}^2,\ldots,c_{jn}^2)\bu_2).
$$
By Central Limit Theorem,
$$
\EE\exp(\bu_1^\top {\rm diag}(c_{j1}^2,\ldots,c_{jn}^2)\bu_2)\to \exp(\zeta^4(\log\log n)/8).%=(\log n)^{\exp(\zeta^4/8)}.
$$
Thus,
$$
\int{f_1^2\over f_0}\to {4\log 2\over 5}\exp[(\zeta^4/8-1)\log\log n].
$$
Therefore, by taking $\zeta$ small enough, we can ensure that any test is powerless in testing $H_0$ against
$$
\bigcup_{\eta\ge 0}\left\{\bmu\in \Theta_{C_n}(\eta^2): \|\bmu\|^2\le {\zeta^2\over 2\sqrt{2}}(n\eta\log\log n)^{2/5}\right\},
$$
which completes the proof.
\end{proof}

\end{document}